\documentclass[acmsmall]{acmart}

\AtBeginDocument{%
  \providecommand\BibTeX{{%
    \normalfont B\kern-0.5em{\scshape i\kern-0.25em b}\kern-0.8em\TeX}}}

\setcopyright{acmcopyright}
\copyrightyear{2018}
\acmYear{2018}
\acmDOI{XXXXXXX.XXXXXXX}

\usepackage{subcaption}
\usepackage{caption}
\newcommand*{\rom}[1]{\expandafter{\romannumeral #1\relax}}
\begin{document}

\title[AIGC in AR Storytelling]{An Exploratory Study on Multi-modal Generative AI in AR Storytelling}

\author{Hyungjun Doh}
\affiliation{%
 \institution{Purdue University}
 \city{West Lafayette}
 \state{Indiana}
 \country{USA}}
\authornote{Marked authors contributed equally to this research.}
\email{hdoh@purdue.edu}
\author{Jingyu Shi}
\affiliation{%
 \institution{Purdue University}
 \city{West Lafayette}
 \state{Indiana}
 \country{USA}}
\authornotemark[1]
\email{shi537@purdue.edu}
\author{Rahul Jain}
\affiliation{%
 \institution{Purdue University}
 \city{West Lafayette}
 \state{Indiana}
 \country{USA}}
\authornotemark[1]
\email{jain348@purdue.edu}
\author{Heesoo Kim}
\affiliation{%
 \institution{Purdue University}
 \city{West Lafayette}
 \state{Indiana}
 \country{USA}}
\email{kim2903@purdue.edu}

\author{Karthik Ramani}
\affiliation{%
 \institution{Purdue University}
 \city{West Lafayette}
 \state{Indiana}
 \country{USA}}
\email{ramani@purdue.edu}

\renewcommand{\shortauthors}{Trovato and Tobin, et al.}



\begin{abstract}
    Storytelling in AR has gained attention due to its multi-modality and interactivity. However, generating multi-modal content for AR storytelling requires expertise and efforts for high-quality conveyance of the narrator's intention. Recently, Generative-AI (GenAI) has shown promising applications in multi-modal content generation. Despite the potential benefit, current research calls for validating the effect of AI-generated content (AIGC) in AR Storytelling. Therefore, we conducted an exploratory study to investigate the utilization of GenAI. Analyzing 223 AR videos, we identified a design space for multi-modal AR Storytelling. Based on the design space, we developed a testbed facilitating multi-modal content generation and atomic elements in AR Storytelling. Through two studies with N=30 experienced storytellers and live presenters, we 1. revealed participants' preferences for modalities, 2. evaluated the interactions with AI to generate content, and 3. assessed the quality of the AIGC for AR Storytelling. We further discussed design considerations for future AR Storytelling with GenAI.
\end{abstract}



\received{20 February 2007}
\received[revised]{12 March 2009}
\received[accepted]{5 June 2009}

\maketitle
\section{Introduction}
Storytelling is the art of conveying a narrative or story to an audience through words, images, sounds, or other forms of content, often with the purposes of entertainment~\cite{miller2019digital}, education~\cite{national2012storytelling}, persuasion~\cite{van2019storytelling}, or preservation of cultural traditions~\cite{palombini2017storytelling}.
With the rapid development of Augmented Reality (AR), Storytelling has expanded onto AR platforms, enabling the possibility of delivering multi-modal content to enhance interactive Storytelling~\cite{stylianidou2020helping, liao2022realitytalk, liu2023visual}.
This evolution has opened the discussion on the creation and effects of multi-modality in the research field of AR Storytelling.

Creating multi-modal content in AR Storytelling poses challenges.
Thorough preparation is necessary for authors to effectively deliver their narratives to an audience, which encompasses not only organizing the story flow but also creating multi-modal content that goes with the flow, to enhance the audience's comprehension and engagement~\cite{liao2022realitytalk}.
However, creating such content often requires mastery of certain software, expertise in visual or auditory design, and time and effort.


To democratize the users or storytellers from the requirement of expertise in content generation, recent work has explored the possibility of deploying Generative Artificial Intelligence (Gen-AI) in AR Storytelling~\cite{han2023design, antony2023id, bensaid2021fairytailor}.
AI-generated content (AIGC) has introduced a wider range of possible modalities into broader research by enabling multi-modal \textbf{functionalities} such as text-to-image~\cite{rombach2022high}, text-to-video~\cite{khachatryan2023text2video}, text-to-audio~\cite{copet2023simple, liu2024classmeta}, and text-to-motion~\cite{shi2025caring}.
Consequently, the boundaries of AR Storytelling are also to be re-established.
The integration of Gen-AI with AR Storytelling is expected to enable the creation of more complex narratives and presentations at a much quicker pace, seamlessly bridging the gap between authors' intention and visual/audio realization~\cite{liu2023visual, shi2023hci}.

Despite the potential benefit of applying AIGC in AR Storytelling, current HCI research has yet to validate the effect of all possible modalities of AIGC in AR Storytelling.
No prior work has provided a systematic validation of the impact of each modality on the core of AR Storytelling, or the design considerations of the necessary interactions for the users to co-create multi-modal content with Gen-AI for AR Storytelling.
To explore the field and locate the pivot point for future research and development, critical questions are yet to be addressed.
Motivated by the promising landscape of Gen-AI for AR Storytelling, we put forward the essential HCI research questions to shed light on future research and development in this field: \textbf{1. How suitable is multi-modal AIGC for augmenting the elements in AR storytelling?} And further, we question \textbf{2. What are the key HCI considerations when designing an AR Storytelling system utilizing the multi-modal power of Gen-AI?}.

To reveal the design space of multi-modal content in AR Storytelling in our study, we have investigated 223 videos from the existing video-edited augmented presentations. 
As a result, we define the design space consisting of the dimensions of \textbf{Elements} of AR Storytelling and the \textbf{Modalities} of augmentations.
We further implemented an authoring platform that leverages Gen-AI models to generate multi-modal content, and an AR interface for live AR Storytelling with the generated augmentations.
With our platform as a testbed, we conducted two studies (each $N=15$) with participants with rich experience in Storytelling and live presentations.
We experimented with five stories, with which the participants either were assigned certain elements to augment with AIGC or freely chose the elements and modalities to augment their AR Storytelling experience.
We present the analysis of the study results which reveal (1) the participants' preferences for modalities to augment elements, (2) qualitative evaluations of the interactions to co-create the content, (3) qualitative evaluations of AIGC in AR Storytelling in terms of expressiveness, immersion, alignment, and exploration, and (4) empirical comparison of AIGC with human-generated content.
Our contributions are four-fold:
\begin{itemize}
    \item A systematic preliminary analysis of 223 augmented storytelling videos and a resulting design space depicting the elements and modalities in AR Storytelling.
    \item Implementation of an assembled storytelling content generator with Gen-AI supporting the creation of animations, images, music, and text, along with an AR interface for storytelling presentations, supporting speech-to-text, displaying corresponding generated content, and hand-interactable features. 
    \item An exploratory study evaluating the impact of multi-modality AIGC regarding its quality, interactions to create, and capability of augmenting elements of AR Storytelling. 
    \item An in-depth discussion providing design recommendations for future investigations within the domain of AIGC in AR storytelling.
\end{itemize}

\section{Related Work}
In this section, we present the landscape of current research.
We divide the related works into two topics: AR Storytelling and Multi-modality in Gen-AI.
\subsection{AR Storytelling} 

Storytelling is the art or practice of a storyteller narrating stories or events, often involving the weaving of characters, settings, and plots, to engage, entertain, educate, or communicate with an audience~\cite{mandelbaum2012storytelling}.
This art or practice is a production co-constructed by two parties - tellers and recipients (or authors and audiences)~\cite{mandelbaum2012storytelling, schegloff1997narrative}.

Recent research in AR Storytelling has demonstrated the suitability of AR as a medium of Storytelling for its presentation and interactivity in situ in multiple modalities~\cite{miller2019digital, calvi2020we}.
Firstly, AR Storytelling can be deployed on multiple platforms for diverse downstream tasks.
For example, Holloway et al.~\cite{holloway2020using} utilize AR Storytelling for educating university students on historical heritage, where the 3D animation of historical or fictional characters of stories is rendered for the users through an AR mobile app.
A similar piece of research by Hirsch et al.~\cite{hirsch2022embedded} demonstrated that users' historical understanding can be improved by embedding our interfaces with AR stories.
Research has also shown the potential of AR Storytelling in diverse domains such as game design~\cite{markouzis2015interactive, vera2016model, raeburn2022developing}, collaborative learning~\cite{park2015storytelling, markouzis2015interactive, yilmaz2017using}, communication~\cite{tyurina2023leveraging, lee2023exploring}, and multimedia industry~\cite{o2021ar, pavlik2013emergence}.

Secondly, AR Storytelling has shown a promising connection between the authors and the audiences and better understanding and expression of objects that are otherwise hard to describe or share~\cite{lee2023exploring}.
Research also emphasizes improving the user experience from each end of AR Storytelling.
For authors, prior works introduce novel designs of how authors can create content for AR Storytelling~\cite{singh2021story, ketchell2019situated, kim2014vision}.
For example, BlocklyXR~\cite{jung2021blocklyxr} introduces an interactive visual programming environment for authoring digital storytelling with 3D models.
Darzentas et al.~\cite{darzentas2018object} combined Internet of Things-inspired approaches and VR/AR
experience technologies, with 3D scanning content creation techniques to create mobile and adaptable AR Storytelling.
For the audience, research focuses on the understanding of the narratives~\cite{gao2022bridging, shin2022effects, csimcsek2023effects, danaei2020comparing, pintar2023invisible, hu2025gesprompt} as well as the immersiveness in the story~\cite{lee2023exploring}.
Our work focuses on the authoring experiences of the storytellers in AR Storytelling, evaluating the content through the authors' satisfaction and preference.

Thirdly, Storytelling in AR consists of multiple possible modalities.
RealityTalk~\cite{liao2022realitytalk} leverages embedded visuals and animation for engaging and expressive storytelling in AR presentations, where users can interactively prompt, move, and manipulate graphical elements through real-time speech and supporting modalities.
Visual Captions~\cite{liu2023visual} enables video conferencing augmented with AI-generated visual captions to enrich verbal storytelling.
FairyTailor~\cite{bensaid2021fairytailor} enables the production of a coherent and creative sequence of text and images for visual Storytelling.
Besides 2D visual content like images and videos, or 3D content like CAD models or avatar animation, modalities such as text~\cite{liao2022realitytalk, markouzis2015interactive, yilmaz2017using} and audio~\cite{zhou2004interactive, indans2019towards, salo2016backend, singh2021story, bauer2019designing} are also utilized in AR Storytelling to enhance the experience.

Despite the variety of modalities available in AR Storytelling serving specific downstream tasks, the authoring of multi-modal content for AR Storytelling can be challenging due to its requirement of mutually exclusive expertise in content creation across diverse modalities.
To this end, researchers in this field have begun to utilize Gen-AI to create multi-modal content for AR Storytelling, enabling the authors to create content of modalities outside of their expertise from those they can easily compose, (e.g. text-to-image~\cite{liu2023visual, petridis2023anglekindling, antony2023id, bensaid2021fairytailor} and image-to-text~\cite{chung2022talebrush, bala2022writing}).
However, prior research has not addressed the impact of multi-modal AIGC on the elements of AR Storytelling.
Our research aims to investigate how multi-modal AIGC can affect the experience of the authors in the pipeline of AR Storytelling.

\subsection{Multi-modality in Gen-AI} 
With the significant development of Gen-AI, researchers have proposed various AI models that generate various modality content, including images, videos, audio, and motion. 

Recently, Gen-AI has achieved significant progress in image generation and manipulation. Researchers have employed Generative Adversarial Networks (GANs) with CLIP~\cite{goodfellow2014generative, karras2021alias, karras2019style}—utilizing image-text representations—for text-driven image manipulation, enabling realistic editing using text~\cite{abdal2022clip2stylegan, gal2022stylegan, mokady2022self, patashnik2021styleclip, duan2024conceptvis, duan2025investigating, duan2025designfromx}. Moreover, with the popularity of diffusion models, several large text-to-image models have been proposed, including Imagen~\cite{saharia2022photorealistic}, DALL-E2~\cite{ramesh2022hierarchical}, GLIDE~\cite{nichol2021glide}, Parti~\cite{yu2022scaling}, CogView2~\cite{ding2022cogview2}, Stable Diffusion~\cite{rombach2022high}, and DreamBooth~\cite{ruiz2023dreambooth}.

In addition to image content, researchers have also investigated video generation using AI.
Prior works utilize the text-to-video method~\cite{singer2022make, ma2023follow, khachatryan2023text2video, kim2020tivgan, chen2023control,shi2023understanding}. 
Recently, video generation models have been implemented based on diffusion models~\cite{ho2022imagen, khachatryan2023text2video, blattmann2023align, kim2020tivgan, chen2023control}. 
Additionally, researchers have explored using different input types, such as image~\cite{ni2023conditional} and motion~\cite{khachatryan2023text2video, ma2023follow}.

Gen-AI has also been utilized for audio generation, employing a variety of models.
Audio generation models have utilized likelihood-based models~\cite{mehri2016samplernn, oord2016wavenet, kalchbrenner2018efficient}, VAE-based models~\cite{peng2020non}, and GAN-based models~\cite{kumar2019melgan, binkowski2019high, yamamoto2020parallel}.
Recently, diffusion models have also been applied to generate audio~\cite{kong2020diffwave}.
Copet et al. introduced MusicGen~\cite{copet2023simple}, a text-to-music model comprised of a single-stage Transformer Language Model (LM).

In the case of human motion generation, the text-to-motion task is used to learn a latent space for language and motion.
JL2P~\cite{ahuja2019language2pose} uses an auto-encoder to learn KIT motion-language dataset~\cite{plappert2016kit}.
T2M employs a VAE~\cite{kingma2013auto} to map a text prompt into a normal distribution in latent space.
Additionally, MotionCLIP~\cite{tevet2022motionclip} addresses the problem of data limitation and enables latent space editing by utilizing CLIP~\cite{radford2021learning}. Researchers also suggest diffusion models for motion generation~\cite{tevet2022human, kim2023flame, zhang2022motiondiffuse}. 

These prior works utilize easily accessible input, enabling users to generate output without requiring any specific expertise.
Text is commonly employed as input to generate content.
In addition to text, prior works have explored the use of various modalities as inputs for Gen-AI, such as image~\cite{ni2023conditional} and motion~\cite{khachatryan2023text2video, ma2023follow}.
This flexibility allows users to generate more diverse and well-aligned content with their intentions, as they are not limited to a single input modality.


    


\section{Design Space of Multi-Modal AR Storytelling}
To investigate the possible modalities in AR Storytelling as well as the elements of Storytelling created through these modalities, we analyzed 223 videos from existing augmented-reality storytelling videos taken from the YouTube platform. 

\begin{figure}[h]
  \centering
    \includegraphics[width=\textwidth]{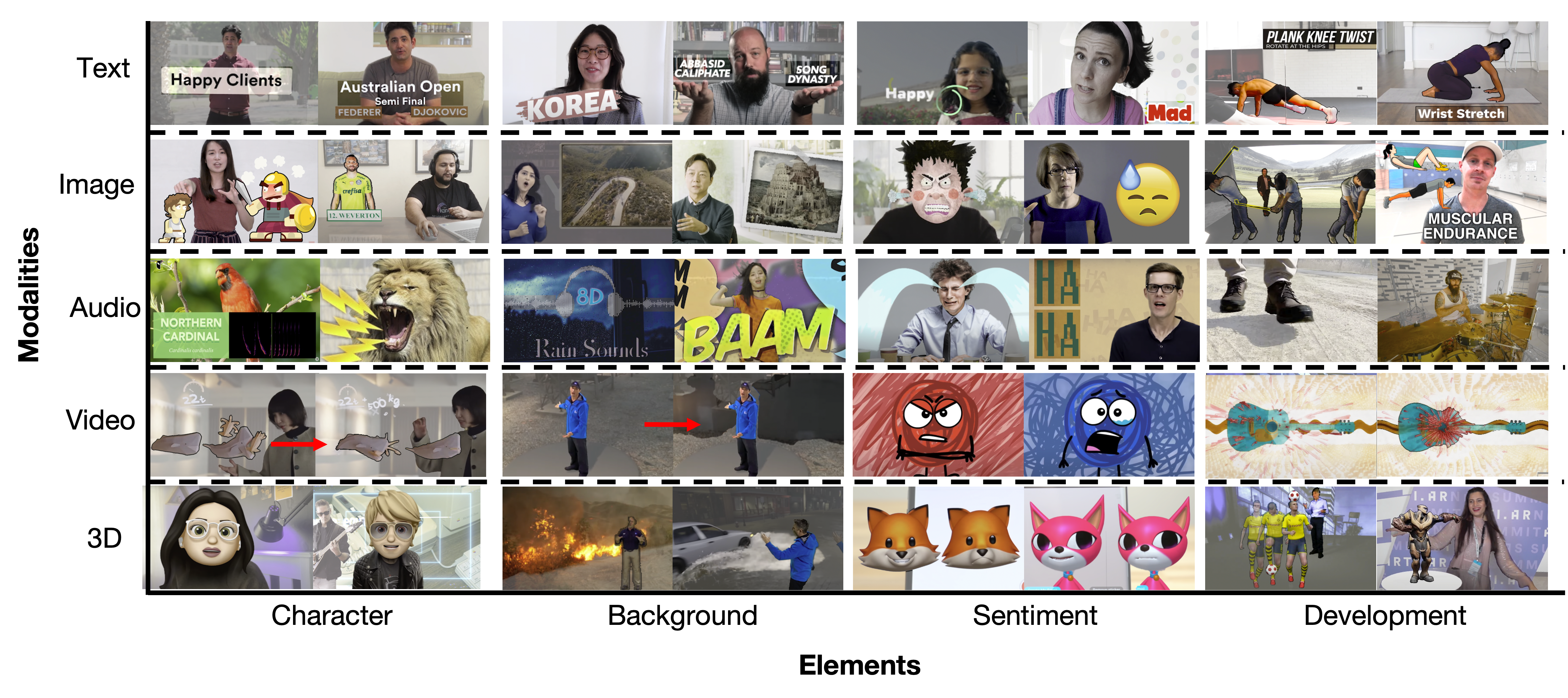}
    \caption{Our Design Space for Multi-modal AR Storytelling: Modalities and Elements. These examples illustrate how a modality augments each element of AR Storytelling, offering a visual summary of our design space as derived from our analysis of the 223 augmented videos.}
    \label{designspace}
\end{figure}

\subsection{Methodology and Ananlysis}

\subsubsection{Identification of Corpus}
To find relevant videos of Storytelling in AR, authors performed a manual search on platforms such as YouTube.
Initially, three authors independently used the keywords "Augmented Reality Storytelling" on YouTube. 
However, this approach yielded few relevant results. 
This was primarily due to two reasons 1) Augmented reality keywords are less used in the AR Storytelling context. 
2) The AR Storytelling videos are often used for different purposes or domains.  
Therefore, we collected videos based on a manual search with keywords of various other domains such as weather, education, tutorial, sports, and news.
We do not claim that our corpus was collected by a systematic search. 
However, we argue that it is sufficient to serve the main goal of systematic analysis to identify common elements and modalities for AR Storytelling.

\subsubsection{Filtering Methodology}
We initially collected 732 videos from YouTube for Storytelling across various domains. 
Then, we conducted a filtering process to exclude the videos that did not possess AR components.
Our inclusion criteria for AR components consist of:
\begin{itemize}
    \item digital information or objects overlaid onto the real-world components in the videos,
    \item audio or visual effects that are not inherently from the storytellers,
    \item auxiliary texts displayed in addition to storytellers' narratives,
    \item animation or dynamic widgets rendered in the videos.
\end{itemize}
Three authors individually watched all the videos and decided on the inclusion or exclusion of the videos.
The inter-rater agreement measure Fleiss' kappa was used during the filtering of the videos.
The mean agreement score of $\kappa$ = 0.76 indicates substantial agreement.
Any conflicts were resolved by discussion among the authors.
After the filtering process, we identified 206 videos. 
Additionally, we added 17 videos based on the author's knowledge from published work in AR Storytelling.
In total, the final corpus contains 223 videos that were used for analysis. 

\subsubsection{Analysis}
To obtain the design space of AR components in Storytelling, we conducted a nuanced analysis of the located corpus.
Notice that the main focus of our analysis was on the content in AR Storytelling and not on the interactions performed with the content.

Firstly, three of the authors individually conducted open coding to analyze a subset of the final corpus ($N=50$) and each summarized a preliminary design space, considering the modalities and elements of AR Storytelling in the videos from each subset.
Secondly, the three authors discussed and decided on a revised design space.
In this step, the purpose of the discussion is to form a clarified definition of each dimension, ensure a clear borderline that differentiates each dimension, and maximize the coverage of the revised dimensions over the three subsets aforementioned.
Thirdly, the authors went through the entire final corpus, individually located each video in the revised design space, and merged to discuss the location of the video.
This procedure ensured each video was reviewed according to a set of consistent criteria.
Upon disagreement about a video, authors merged and discussed to decide whether existing dimensions were appropriate or add/remove dimensions otherwise.
Additionally, our main finding on analyzing the videos suggests that speech modality (speech by the storytellers, which is a necessity) triggers other modalities such as text, visuals, or audio sounds relevant to the narration in AR Storytelling.
Therefore, we used the narrated content as our primary modality to design our authoring system which then triggers other modalities in the AR Storytelling.

\subsection{Design Space}

To this end, we identified the design space of multi-modal AR Storytelling, covering the discussion over possible modalities of the AR components and the elements of AR Storytelling that can be represented and expressed by the modalities (as shown in ~\autoref{designspace}). 

\subsubsection{Modality}
We identified 5 different types of modalities being used in the current field of AR Storytelling:
\begin{itemize}
    \item \textbf{\textit{Text}}: In AR Storytelling, the text is overlaid in a real environment to provide background stories characters dialogue, or additional plot details enhancing the overall storytelling experience.
    Texts are usually used for detailed information delivery such as instructions, context, or any other details, and provide user guidance such as navigating or prompts.
    \item \textbf{\textit{Audio}}: Many AR Storytelling scenarios involve background sounds, music, or ambient noises, which contribute to the mood and atmosphere of storytelling.
    Character voices and dialogues make storytelling experiences more immersive.
    Realistic sounds are associated with virtual objects making the experience more engaging.
    Audio cues are used as feedback for user interactions.
    For example, a sound effect confirms a successful button click.   
    \item \textbf{\textit{Image}}: Images are used to convey key visual elements of the story, such as the appearance of characters, the vision of scenes, or the composition of objects.
    Visual elements add depth to the narrative making it more memorable for the users.
    Images are used to represent characters, and interactive elements and provide visual feedback.
    Visual images also have a powerful impact on rendering emotions.
    It enhances the emotional resonance of the narrative.
    Visual imagery is also used for aesthetics and designs of AR Storytelling. 
    \item \textbf{\textit{Video}}: Videos can convey dynamic and moving content, allowing for a more fluid and cinematic Storytelling experience.
    Animation and motion make characters or virtual objects move.
    Character animations and temporal narrations convey sequences of actions, and changes in the storyline, or illustrate the development of the plot.
    \item \textbf{\textit{3D}}: 3D models are used as characters, objects or scenes.
    3D models add more dimensions and viewpoints to visual AR content, e.g. 3D Avatars, 3D texts, 3D scenes, and 3D animations.
\end{itemize}

\subsubsection{Elements}
We identified four different atomic elements essential to telling a story in AR.
These atomic elements, namely Character, Background, Sentiment, and Motion, compose higher-level elements of a complete story such as Conflict, Plot, and Message~\cite{fog2005storytelling}, and eventually serve the highest-level goal of the story - to make the audience understand~\cite{schegloff1997narrative}.

\begin{itemize}
    \item \textbf{\textit{Character}}: A Character refers to a person, an entity, or a creature that inhabits the story.
    It can be a human, an animal, an object, or even an abstract concept, depending on the nature of the narrative.
    A Character can be elicited through its traits, roles, appearance, or other attributes that help shape its figure in the story.
    \item \textbf{\textit{Background}}: This element encompasses the setting or environment where the story unfolds.
    On the one hand, it can be a physical location where the story happens such as a castle, a forest, or a beach.
    On the other hand, it can be depicted as a virtual context or an abstract setting, that contributes to the cultural or historical context of the story, e.g. London in the eighteenth century or a world with no elves.
    The atomic background element solely depicts the ``where'' and ``when'' the story happens.
    \item \textbf{\textit{Sentiment}}: Sentiment refers to the emotional tone, mood, or atmosphere conveyed through the story experience.
    In Storytelling, characters (usually humanoid ones) can display their Sentiments that add context and reasoning to the plots, shape the figure of the characters, or stir up sentimental resonance with the audience, e.g. ``Jack is sad due to his loss''.
    Furthermore, the narratives can also display Sentiments that render the atmosphere of the story without being derived from the characters, e.g. ``The devastating smell resonates in the air''.
    \item \textbf{\textit{Development}}:  Development includes dynamic movements and temporal progress of the Backgrounds, Characters, or Sentiments.
    In Storytelling, Development depicts the time arc of the story, which emphasizes the progression or change in a significant transition along the timeline, e.g. ``Lucy turned twenty-two and graduated from college''.
    The Development element also involves motion, action, or movement that focuses on the behaviors or changes that do not emphasize time, e.g. ``The cup keeps spinning" and "Andrew is a lot happier now''.
\end{itemize}
\section{Exploratory Study}
\subsection{Study Setup: Testbed Implementation}
To investigate our research questions, we conducted an exploratory study to evaluate the effect of multi-modal AIGC in the preparation and presentation stages of AR Storytelling. 
We thus developed a multi-modality content generator and AR Storytelling platform as the testbed for our study.

\begin{figure}[htp]
  \centering
    \includegraphics[width=0.65\textwidth]{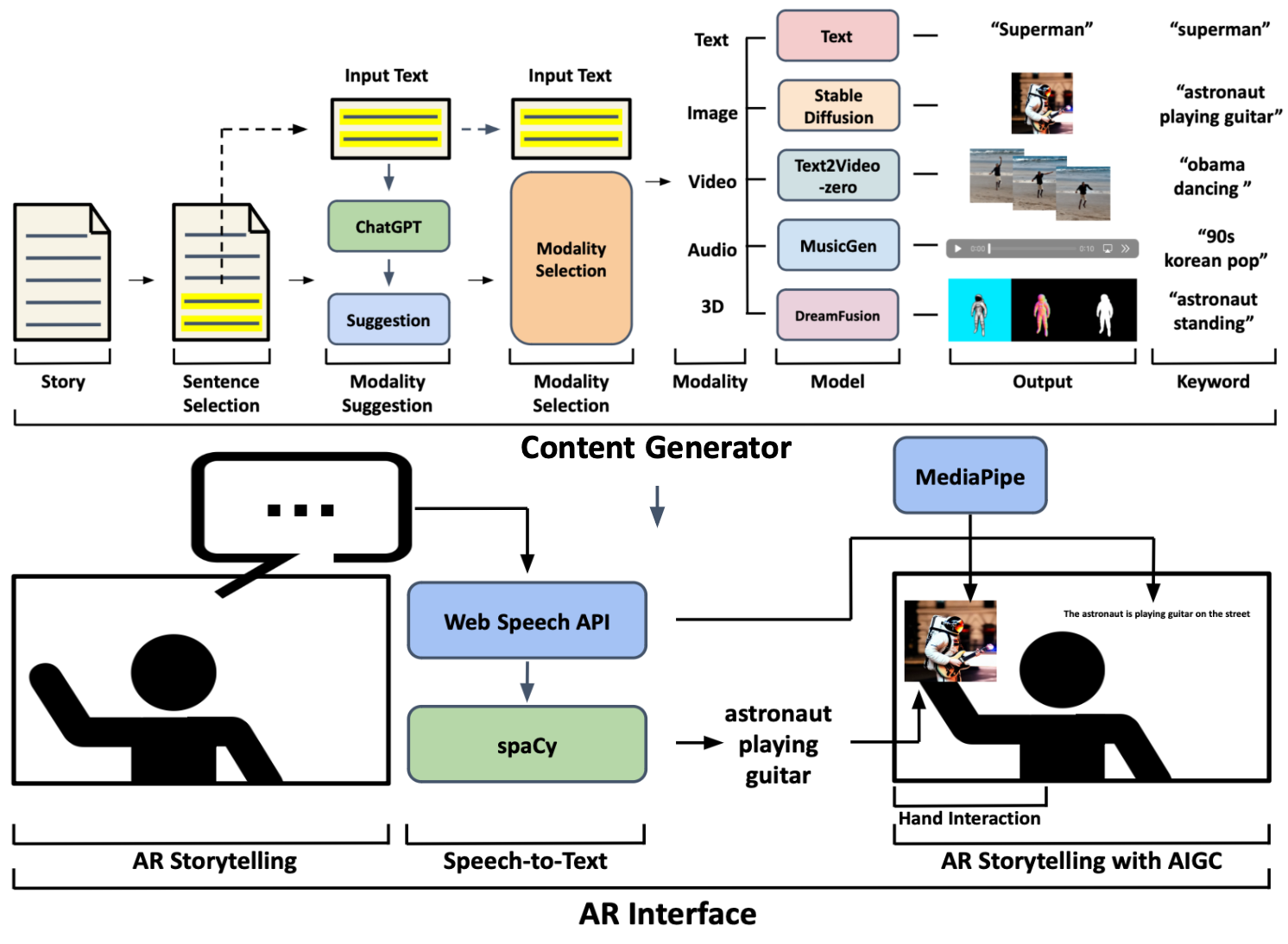}
    \caption{The testbed workflow. 1) \textbf{Content Generator interface: } The user employs the content generator to create AIGC for AR Storytelling, supporting five modalities for the selected sentence. 2) \textbf{AR interface: } The user can view the text corresponding to spoken words. Based on the transferred speech text, the user can interact with AIGC using hand.}
    \label{workflow}
\end{figure}

The testbed contains two separate interfaces. 
1) \textbf{Content Generator interface:} the author can create the content for the AR Storytelling using AI according to textual narratives as shown in ~\autoref{testbed}-(a).
2) \textbf{AR interface:} This AR interface allows users to interact and place the generated content at the desired location and present the entire story augmented with multi-modal content as shown in ~\autoref{testbed}-(b).

The testbed is designed with the following specified considerations: 
1) The testbed should generate multi-modal AR components, including text, audio, image, video, and 3D, with a given text narrator. 
2) The testbed should be able to generate content with various forms of sentence structure.
3) The testbed should allow the users to decide the modality of the AIGC based on their preference.
4) Through iterations with the users, the testbed should be able to create content aligning with the author's intention. 
5) The testbed should allow AR Storytelling with generated content, supporting simple spatial AR interaction built with hand gestures, and speech-based interactions. 

\subsubsection{Multi-modal Content Generator}
Our multi-modal content generator is an interface empowered by multiple Gen-AI models based on textual prompts.
For video generation, our interface employs the Motion-Diffusion-Model (MDM)~\cite{tevet2022human}. 
This model takes text as input and generates detailed character motion (For example, generating a human motion of skipping rope).  
To extract the verb and noun from the input sentence, we leverage spaCy~\cite{honnibal2020spacy}, a Natural Language Processing (NLP) Python library. 
Then based on the generated motion, our system uses Text2Video-Zero~\cite{khachatryan2023text2video} to create a video with the described character and background. 
For image generation, we utilize Stable Diffusion~\cite{rombach2022high}.
The testbed generates the audio with MusicGen~\cite{copet2023simple} which generates audio using text prompts.
To generate 3D content, we used text-to-3D Gen-AI model~\cite{poole2022dreamfusion}.

~\autoref{testbed}-(a) illustrates the front-end interface of our multi-modality content generator testbed.
The front-end communicates with the back-end using FastAPI ~\cite{fastapi}, a web framework for building RESTful APIs~\cite{restful} in Python. 
The generator comprises three main sections:
(1) \textbf{Story input}: The textual input section imports the text file containing the narratives.
Users can select the sentence in this section by highlighting the portion, which they want to augment in the storytelling process and (2) \textbf{Output}: Based on the users' selection of modality, the output section generates and displays the corresponding AR component.
The output is saved, corresponding to the highlighted sentence, which is a trigger during the storytelling process. \\

\subsubsection{AR Interface}
\begin{figure}[htp]
  \centering
    \includegraphics[width=0.9\textwidth]{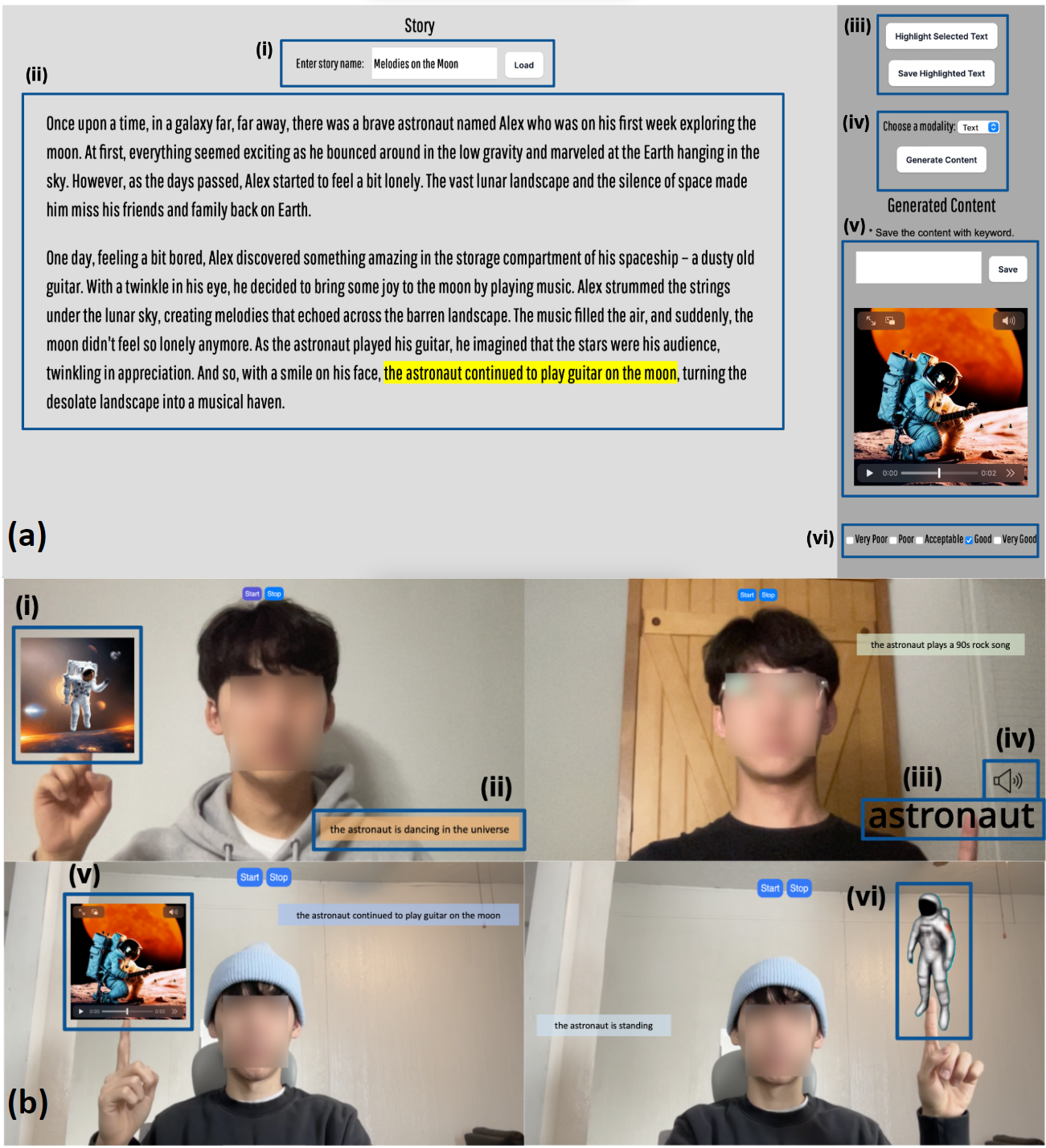}
    \caption{The testbed. \textbf{(a)} The Multi-Modal Content Generator interface. (a-\rom{1}) The textual input section. The user can import a story text file for storytelling by entering the story title. (a-\rom{2}) The loaded story. Where the user can see the loaded story and select a sentence for augmentation by dragging it with the mouse cursor. (a-\rom{3}) The highlight (top) and save (bottom) buttons. The top button highlights the selected portion with yellow, indicating to the user that this part will be generated. The bottom button saves the highlighted text to the backend for output generation. (a-\rom{4}) The modality selection. The user can choose the desired modality for content generation. (a-\rom{5}) The output. The testbed displays the generated content, and the user can save it using a unique keyword. This keyword acts as a trigger during the storytelling process. (a-\rom{6}) The quality evaluation of the content by the user. \textbf{(b)} The AR interface. (b-\rom{1}) Image. The user can interact with the AIGC using hand landmarks. The content appears on the tip of the user's index finger. (b-\rom{2}) The Speech-to-Text box. The interface displays the narrator's words, serving as a trigger for the corresponding content. (b-\rom{3}) Text. This modality shows the main keyword and detailed information. (b-\rom{4}) Audio. The icon indicates the corresponding audio is playing. (b-\rom{5}) Video. (b-\rom{6}) 3D Content.}
    \label{testbed}
\end{figure}

~\autoref{testbed}-(b) shows the interface of our AR Storytelling platform, testing the AIGC in real AR Storytelling scenarios to observe how it impacts the storytelling experience.
The AR Storytelling interface incorporates MediaPipe ~\cite{mediapipe} for Hand Landmark detection to support hand interaction to locate the generated content.
Additionally, it employs the Web Speech API~\cite{mozilla} to showcase the narrator's words on display and trigger the generated content. Users can convey their stories in AR using AIGC with these features. 

\subsection{Study Design}
We conducted two studies to evaluate the effect of multi-modal AIGC in AR Storytelling.
Study 1 was performed in a controlled setting to evaluate which modality is best suited for what type of elements of AR Storytelling according to the authors' preference.
Study 2 was performed to evaluate users' interactions in generating content for AR Storytelling.
Furthermore, we aimed to gain insights into the deployment of multi-modal AIGC in AR Storytelling through subject participant feedback for both studies.

\subsubsection{Participants}
We invited 30 participants (13 female; and 17 male) via emails, mutual connections, and posters from a university’s graduate and undergraduate programs.
The user's age ranges between 18 to 31 with a mean of 25 years. 
All of the users had prior knowledge of Storytelling in the form of presentations, oral speeches, theater performances, etc.
All participants had prior experience in live public storytelling such as stand-up comedy shows, presentations to a crowd audience, or being interviewed.
24 out of 30 participants had some experience with AR/VR interfaces.
The participants were randomly divided into two groups, each participating in one of the two studies.
After the arrival of users, a 5-minute brief introduction was provided about the study. Users agreed to perform the study and signed a consent form.
A step-by-step instruction was given to the participants on the usage of the interface and the overall goal of the study
Then, the participants were given sufficient time to get familiar with the interface up to 15 minutes.
Both studies took 1.5 hours in a closed environment and each participant was compensated with a 15 USD e-gift card. 

\subsubsection{Study 1 - Designated Augmentation}
In this study, we aim to investigate the relationship between AR Storytelling elements and modality.
We designed 5 different narratives of stories, each with around 20 atomic elements priorly highlighted by us in the interface.
The narratives are shown in ~\autoref{apdx:stories} and the pre-determined elements statistics are shown in ~\autoref{fig:element_dist}.
For each element previously highlighted in the interface, the participants are asked to select at least one modality of augmentation that they think helps with storytelling the most.

Participants are allowed to preview all modalities of the generated content on the generator interface.
They are also allowed to prepare or demonstrate storytelling on the AR interface with the generated content.
Each participant was asked to complete the augmentation and practice of AR Storytelling of all 5 stories.
Upon completion of selecting the modalities, creating the augmentation, and practicing the AR Storytelling with the AIGC, the participants completed a Likert-scale questionnaire and received a semi-structured interview about the experience.

\subsubsection{Study 2 - Freestyle Augmentation}
In this study, we aim to investigate participants' preference for modalities to augment their AR Storytelling experience and the interaction between the participants and the AI models in the process of content generation.
We utilized the same 5 stories as in Study 1.
Differently, participants were not given specific elements from the narratives to augment.
Instead, participants were asked to augment their storytelling experience according to their preference, by selecting and highlighting elements from the narratives and augmenting them at will.
The content generation process, the preview process, and the practice process were all the same as those in Study 1, with an additional request to give ratings on a scale of 5 for the quality of each augmented element.
Participants were also asked to finish augmenting and practicing AR Storytelling with all 5 stories.
Upon completion of selecting the modalities, creating the augmentation, and practicing the AR Storytelling with the AIGC, the participants completed a similar Likert-scale questionnaire and received a semi-structured interview about the experience.

\subsubsection{Data Collection}
As described in the previous subsubsections, we collected the following data:
\begin{enumerate}
    \item 5-point Likert-scale questionnaires indicating the participants' subjective ratings of the creation and practicing of the AR Storytelling experience,
    \item post-study semi-structured interviews,
    \item video recordings during the entire for each participant.
\end{enumerate}
When designing the Likert-scale questionnaires, we took into consideration the Expressiveness, Immersiveness, Exploration, and Alignment in the co-creation process with AI, as suggested in the Mixed-Initiative Creativity Support Index (MICSI)~\cite{lawton2023drawing}.
We only considered a subset of the factors in MICSI to be evaluated, given that our testbed is not an implementation of a systematic application, and thus systematic evaluation metrics such as usability do not contribute to the findings.
In addition to the above, we collected the participants' preferences for modalities over the pre-determined elements in Study 1.
In Study 2, we collected the participants' ratings of AIGC while creating content.
Also, we reviewed the recordings to analyze their failed attempts to augment with some modalities (where participants generated the augmentations but did not apply them later in AR Storytelling practice).

\section{Results}
\label{sec:results}
In this part, we present our results from the study.
Then we summarize and analyze the results according to the interview feedback and our observations.

\subsection{Preferences for Modalities to Augment}

\begin{figure}[htp]
     \centering
\begin{minipage}[t]{0.4\textwidth}
    \centering
    \includegraphics[width=\linewidth]{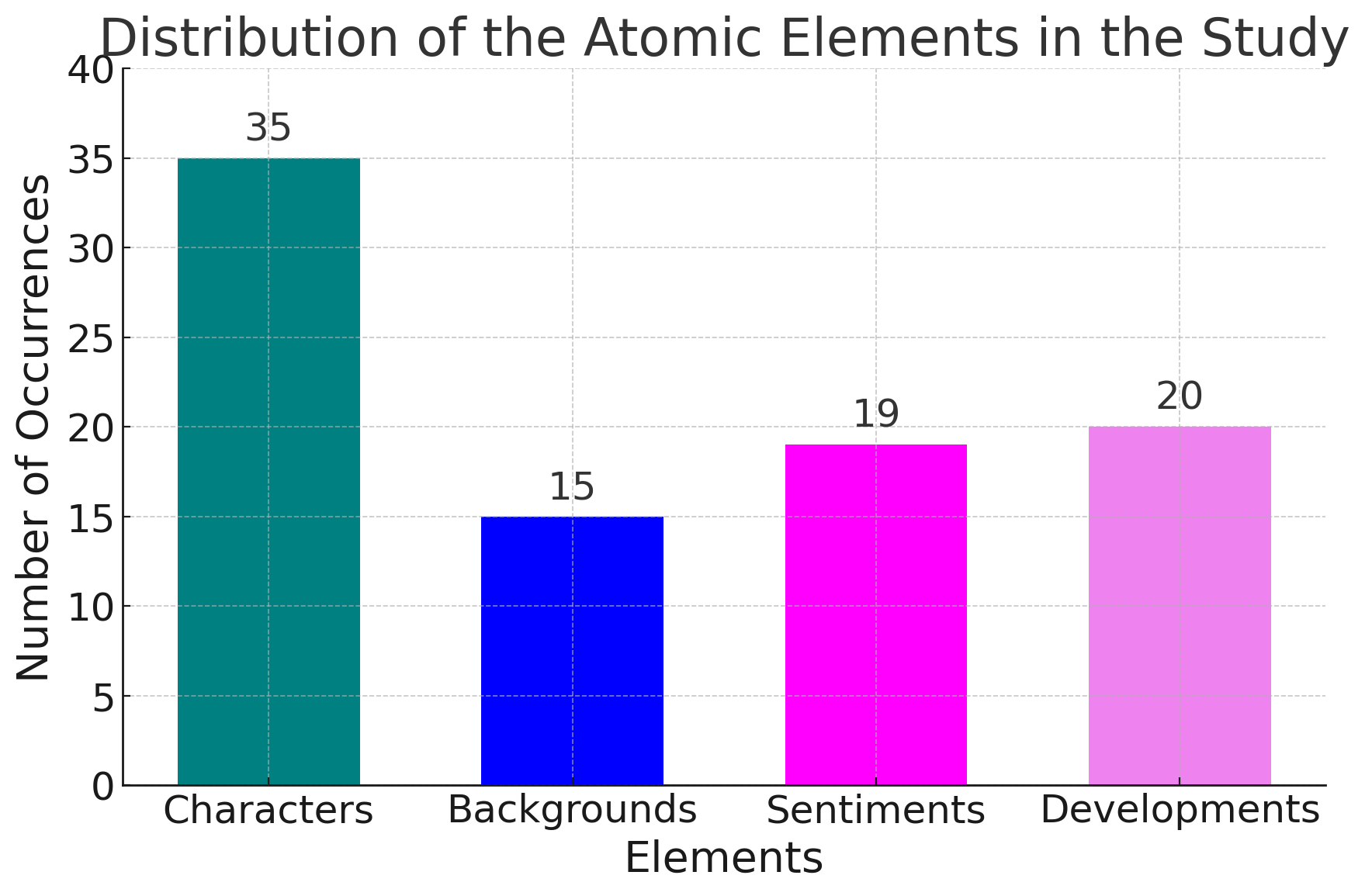}
    \caption{The distribution of the four elements that emerge in the stories we used for the study}
    \label{fig:element_dist}
\end{minipage}
~
\begin{minipage}[t]{0.5\textwidth}
    \centering
    \includegraphics[width=\linewidth]{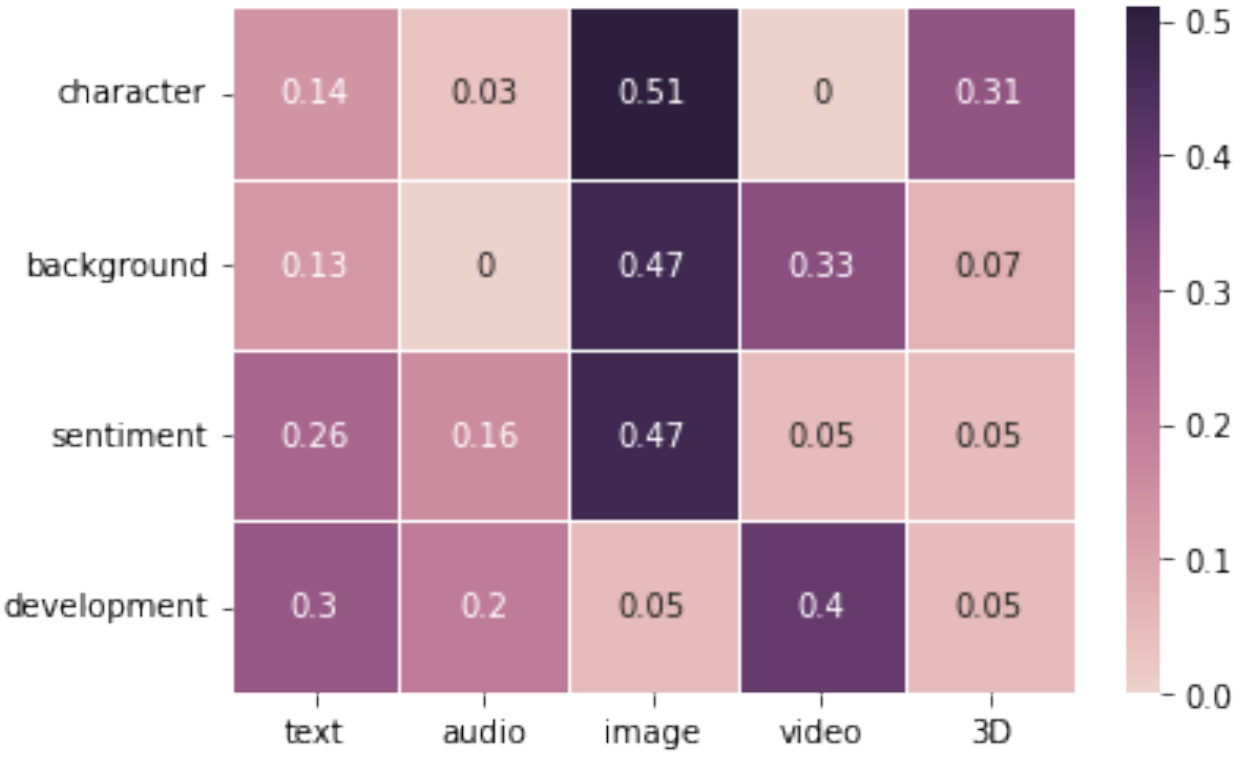}
    \caption{Results for Preference of modality } 
    \label{fig:preference_study1}
\end{minipage}
\end{figure}

We report our findings in ~\autoref{fig:preference_study1} which summarizes which modality should be used for what elements of AR Storytelling in Study 1.
Characters are highly preferred by images (51\%).
\textit{``In AR stories, when we see pictures of characters, it feels like they're right there with us, making the story come alive.'' (P4)}.
3D (31\%) is also used to represent characters while other modalities are slightly preferred.
\textit{``I use 3D characters often in cases for showing the action of the character'' (P6)}.
For background, image (47\%) and video (33\%) are best suited.
\textit{``I prefer 2D visuals in the background'' (P9).}
The favorite modality to express sentiment is image (47\%) followed by text (26\%).
\textit{``I think emotions and feelings can best be expressed by showing image'' (P1)}.
Text (30\%) and video (40\%) were preferred by participants for development elements and then audio.
\textit{``I feel like the video was easy for me to show change'' (P14).}


\begin{figure}[htp]
    \centering
    \includegraphics[width=0.6\linewidth]{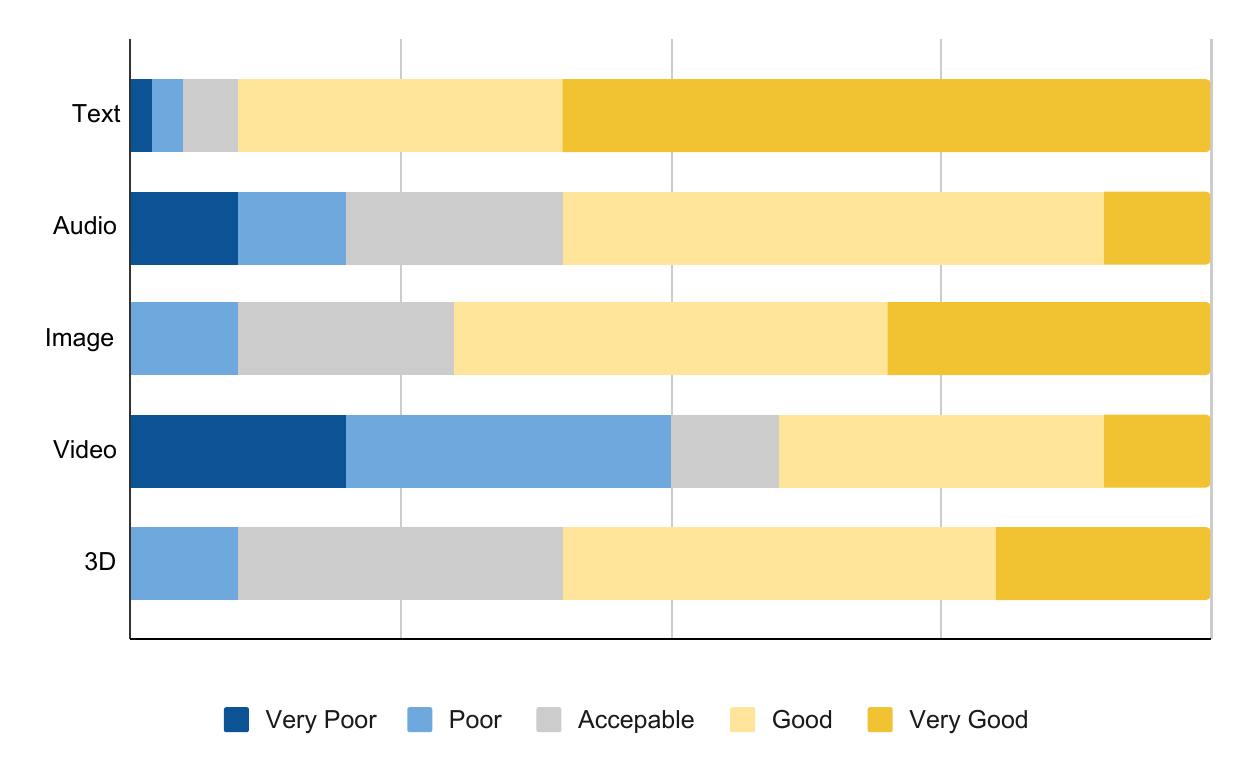}
    \caption{The participants' ratings of the overall quality of each modality.}
    \label{fig:preference_study2}
\end{figure}

\subsection{Quality of AIGC in AR Storytelling}

We evaluated the quality of the AIGC during the study 2.
The results are summarized in ~\autoref{fig:preference_study2}. 
Participants have appreciated the quality of the text (AVG=4.43, SD=0.87).
\textit{``I think the text is best because it was clear and easy to read during AR presentation and are very much relevant to the story ''(P23)}. 
Participants had neutral opinions on the quality of the generated audio from the text (AVG=3.4, SD=1.11).
\textit{``Most of the time the sound I was hearing for my content was suitable'' (P17)}
Participants agree that the images were relevant and good quality (AVG=3.9, SD=0.94){\textit{``The images looked so real as if someone has taken in the real world and were so much similar to the story.''} (P19)}.
Participants were not satisfied with the quality and faced difficulty in understanding the content of the video ((AVG=2.8, SD=1.13)).
\textit{``I was not able to understand what was the action being performed in the video''(P27)}.
Many participants acknowledge the 3D content was useful for their story (AVG=3.7, SD=0.90). \textit{``I cannot imagine I can get 3D models of astronaut with a telescope so good''(P25)}


\subsection{Interactions with AI to Generate Augmentation}

\begin{figure*}[htp]
  \centering
  \begin{minipage}{.45\textwidth}
    \centering
    \includegraphics[width=\linewidth]{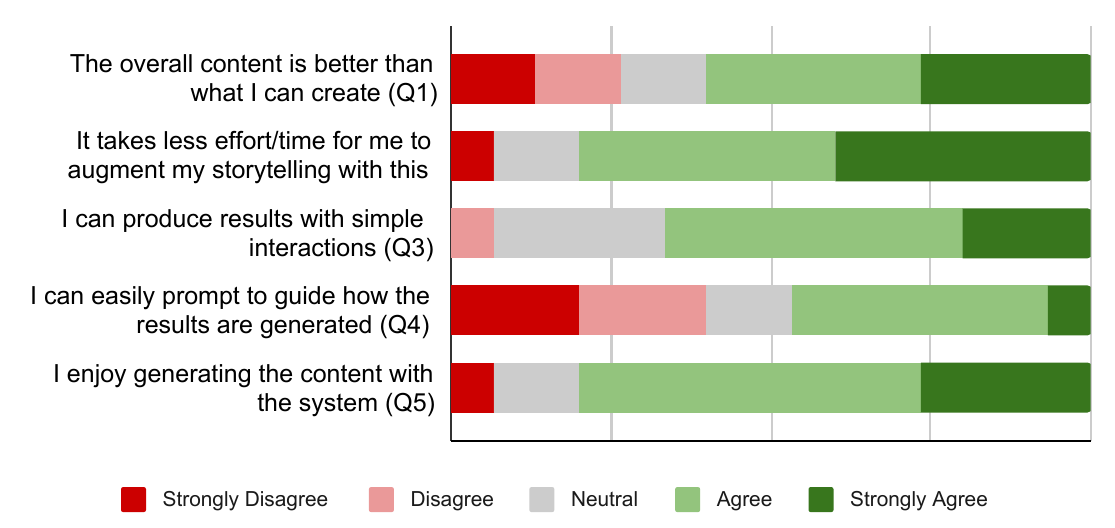}
    \caption{Results of the Likert-scale questionnaires on the interactions with the testbed to generate augmentations - Study 1}
    \label{fig:Study1_interaction}
  \end{minipage}%
  \quad
  \begin{minipage}{.45\textwidth}
    \centering
    \includegraphics[width=\linewidth]{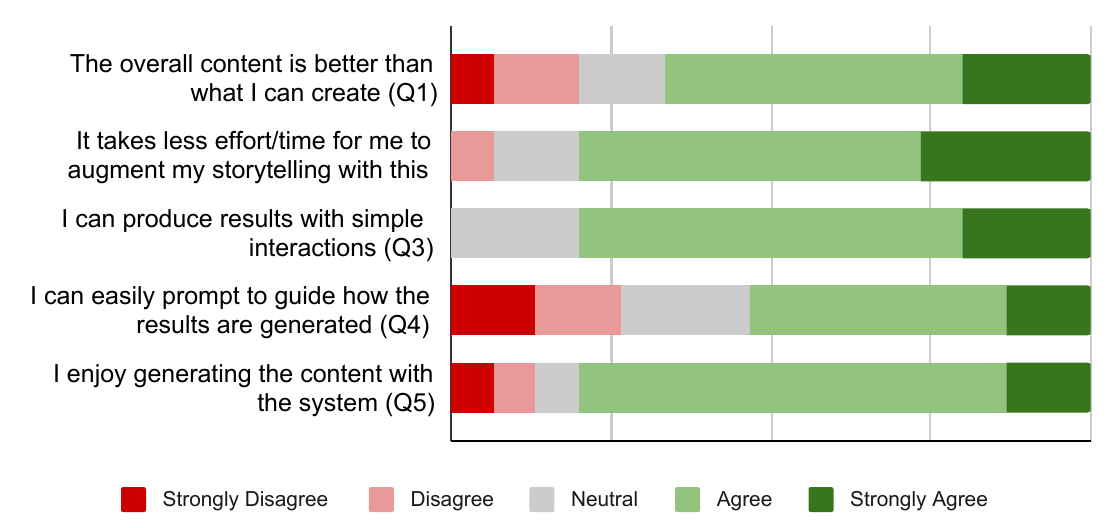}
    \caption{Results of the Likert-scale questionnaires on the interactions with the testbed to generate augmentations - Study 2}
    \label{fig:study2_interaction}
  \end{minipage}
\end{figure*}

We qualitatively evaluated the interactions of the participants with the testbed in generating the content during both Study 1 and Study 2 as shown in ~\autoref{fig:Study1_interaction} and ~\autoref{fig:study2_interaction}.
The participants acknowledge that the final AR Story with the generated content is usable and better than what they can create on their own ((Q1) Study 1: AVG=3.47, SD=1.35; Study 2: AVG=3.60, SD=1.14).
\textit{``This is actually good, I would like to use such systems to generate content for my presentations in the future.'' (P9)}.
Most of the participants were able to generate relevant content with less time/effort ((Q2) Study 1: AVG= 4.06, SD = 1.06; Study 2: AVG= 4, SD = 0.86).
\textit{``In my opinion, this is one of the fastest software I have ever used to make content.'' (P24)}
Participants found that the interaction with the testbed is simple and easy to generate content ((Q3) Study 1: AVG= 3.83, SD=0.82; Study 2:AVG= 4.1, SD=0.63 ).
\textit{``It is definitely not at all a complicated system to use to get some results'' (P5)}.
Participants found it difficult to guide the generation to the desired output using just textual prompts ((Q4) Study 1: AVG=2.93, SD=1.28, Study 2: AVG=3.26, SD=1.23). \textit{``Every time I entered new text, the output looked entirely different and was sometimes not usable.'' (P8)}
Participants enjoyed the generation process of the content using the testbed ((Q5) Study 1: AVG=3.91, SD=0.95; Study 2: AVG=3.73, SD=0.99). \textit{``The system is fun to use, as it gives me some novel options to try for my presentations.'' (P13)}

Additionally, there is no significant difference in the participants' ratings of the interactions between the two studies. Shapiro-Wilk tests on the ratings showed no normally distributed feedback from either group(Q1:$p_{sw,1}=0.030$, $p_{sw,2}=0.036$;
Q2:$p_{sw,1}=0.011$, $p_{sw,2}=0.001$;
Q3:$p_{sw,1}=0.003$, $p_{sw,2}=0.050$;
Q4:$p_{sw,1}=0.072$, $p_{sw,2}=0.037$;
Q5:$p_{sw,1}=0.0005$, $p_{sw,2}=0.001$.), and Mann-Whitney U tests showed no significant difference between groups(Q1:$p_{mwu}=0.920$, $Z_{mwu}=0.103$;
Q2:$p_{mwu}=0.645$, $Z_{mwu}=-0.456$;
Q3:$p_{mwu}=0.589$, $Z_{mwu}=0.539$;
Q4:$p_{mwu}=0.535$, $Z_{mwu}=0.622$;
Q5:$p_{mwu}=0.603$, $Z_{mwu}=-0.518$.
).
This suggests that freestyle augmentation and designated augmentation have no significant difference in the interaction experiences of the participants.



        
\subsection{The Suitability of AIGC in AR Storytelling}

\begin{figure*}[htp]
  \centering
  \begin{minipage}{.45\textwidth}
    \centering
    \includegraphics[width=\linewidth]{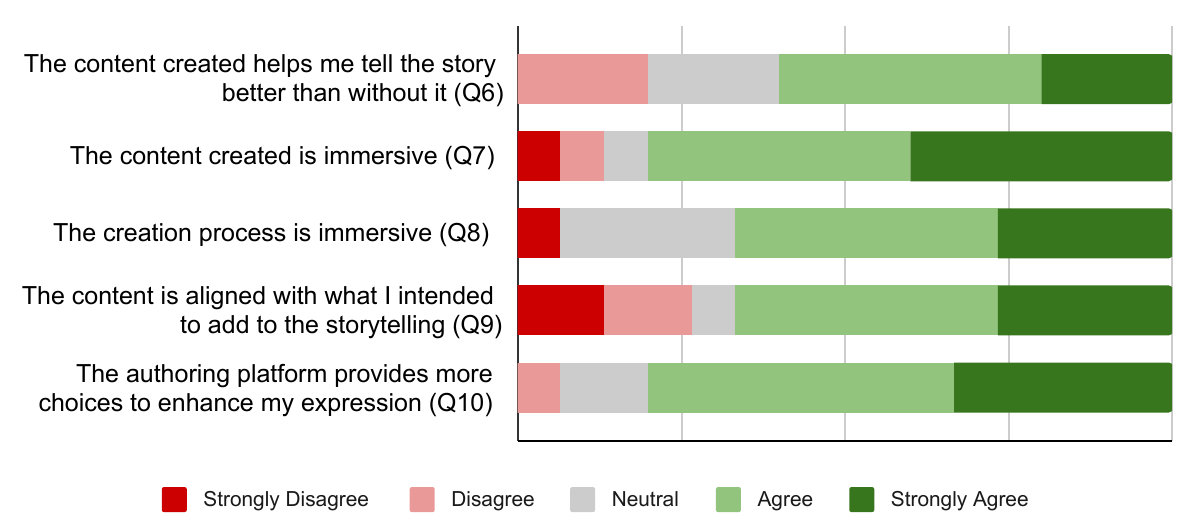}
    \caption{Results of the Likert-scale questionnaires on Suitability - Study 1}
    \label{fig:study1_quality}
  \end{minipage}%
  \quad
  \begin{minipage}{.45\textwidth}
    \centering
    \includegraphics[width=\linewidth]{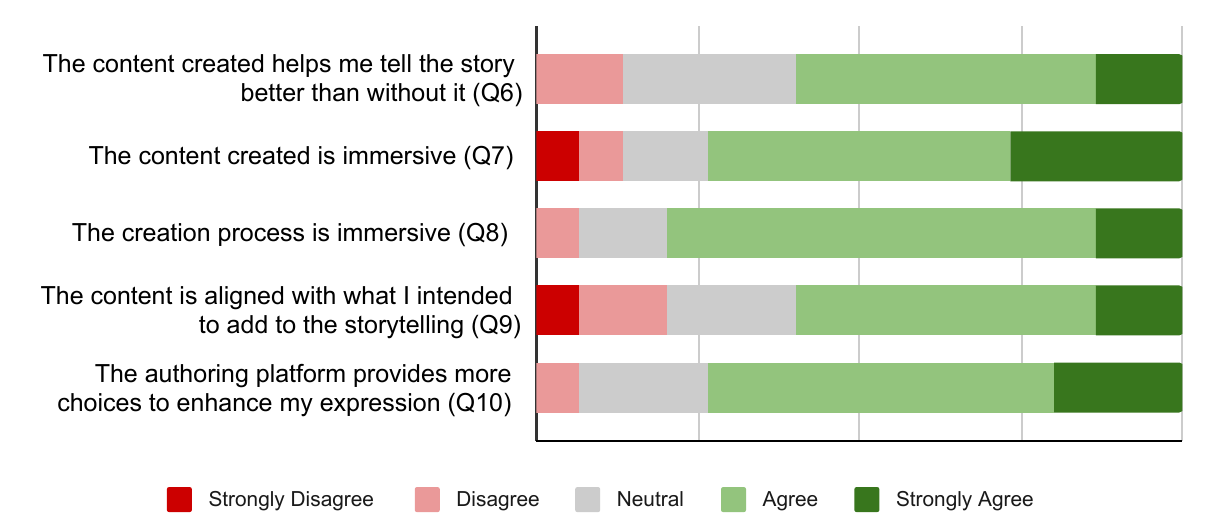}
    \caption{Results of the Likert-scale questionnaires on Suitability - Study 2}
    \label{fig:study2_quality}
  \end{minipage}
\end{figure*}

We qualitatively evaluated the usefulness of the content in both Study 1 and Study 2 as shown in ~\autoref{fig:study1_quality} and ~\autoref{fig:study2_quality}. 
\textbf{\textit{Expression}}: participants found that the content created is useful and able to tell the story better than without using it {((Q6) Study 1: AVG=3.60, SD= 1.01; Study 2: AVG=3.60, SD=0.87)}. \textit{``I think additional generated content is good in narrating the story better than without using any content.'' (P2)}
\textbf{\textit{Immersion}}: participants found the that content created is immersive {((Q7) Study1 : AVG=4.03, SD=1.15; Study 2: AVG=3.80, SD=1.10)}. \textit{``I just think that the content will increase audience engagement and I feel like the audience will better connect.'' (P15)} 
Participants think positively that the content creation process is also immersive ((Q8) Study 1: AVG=3.81, SD= 1.04; Study 2: AVG=3.87, SD=0.71). \textit{``I was so deeply involved with the process that I have to say this is very much authentic.'' (P17)}
\textbf{\textit{Alignment}}: Participants found that the content is aligned with their intentions and the content effectively conveys the narrative {((Q9) Study 1: AVG=3.53, SD = 1.35; Study 2: AVG=3.46, SD = 1.08)}. \textit{``I feel like the content was pretty much the same as what I wanted. I was surprised though that I can get what I want easily.'' (P10)}
\textbf{\textit{Exploration}}: participants reported positively that the authoring platform provides more choices to enhance my expression {((Q10) Study 1: AVG= 4.06, SD= 0.85; Study 2: AVG= 3.86, SD= 0.80)}. \textit{``I wonder how I am able to get so many variations of the content with just text.'' (P3)}

Additionally, there is no significant difference in the participants' ratings of the interactions between the two studies. Shapiro-Wilk tests on the ratings showed no normally distributed feedback from either group (Q6:$p_{sw,1}=0.049$, $p_{sw,2}=0.041$;
Q7:$p_{sw,1}=0.012$, $p_{sw,2}=0.002$;
Q8:$p_{sw,1}=0.002$, $p_{sw,2}=0.017$;
Q9:$p_{sw,1}=0.057$, $p_{sw,2}=0.013$;
Q10:$p_{sw,1}=0.024$, $p_{sw,2}=0.012$.), and Mann-Whitney U tests showed no significant difference between groups(Q6:$p_{mwu}=0.9681$, $Z_{mwu}=0.04117$;
Q7:$p_{mwu}=0.53526$, $Z_{mwu}=-0.62217$;
Q8:$p_{mwu}=0.645$, $Z_{mwu}=-0.456$;
Q9:$p_{mwu}=0.69654$, $Z_{mwu}=-0.39404$;
Q10:$p_{mwu}=0.50926$, $Z_{mwu}=-0.66365$.
).
This suggests that freestyle augmentation and designated augmentation have no significant difference in the participants' ratings of the suitability of AIGC in their Storytelling experience.



\subsection{Result Summary and Analysis}

\subsubsection{Suitability of modality}
It was evident that images were highly preferred for character, background, and sentiment. This is because of progress in algorithmic advancement in image generation as compared to that in other modalities.
3D content was mostly used for depicting characters.
This is mostly because 3D models are enough to introduce the rough figure of the characters while the background, scenes, or minute details such as the expression, development, or appearance of the character are not easily and accurately rendered in 3D models.  
Though some participants highly preferred the text regarding its quality, others felt that the text did not add details.
\textit{``I mean I have used text for almost everything but the text is not that engaging as compared to visuals or audio''. (P13)}
The quality of the video was not good enough so participants only used video at a big scale in the development element where details are safe to ignore, while for small details in development, participants preferred text and audio (P4, P2, P9, P13).
\textit{``We used video for development mostly for expressing bigger ideas because of its poor quality. I feel like small details are not good'' (P9).}
Furthermore, the quality of the videos was reported as the major reason for the misalignment between the content and the intention of the participants.
As the participants' speech narration made up the majority of the audio content in the storytelling, generated audio was less used for all the elements and mostly functions as a supplement to describe some of the elements.
Audio is also used for showing emotions, and sometimes the temporal progress of the scenes (P1, P6, P15).


\subsubsection{Empirical comparison to human-generated content}
Many participants appreciated that the overall AR Storytelling with AIGC was useful.
Some of the participants felt the content was as good as human-generated content while some felt that the quality of the content limits their freedom to use some modality for some element.
The most negative feedback we received from the participants was on the quality of the videos.
While participants reported that the generated videos were far from being as good as human-recorded or generated ones, we argue that this is due to the limitations in the algorithmic development in content generation rather than the suitability of the video (or others) as a modality for augmenting storytelling.

Interestingly, participants reported that they could tell the difference between the human-generated content from the AIGC according to the interview results (P1-4, P6, P9-P12, P16-23, P25).
However, most of these participants commented that it does not affect the quality of their storytelling, since it is not the story itself but just the augmentation.
\textit{``I can tell that the image is definitely not drawn by human[s] but I kind of think that it is still related to my story.'' (P5)}
Yet, some commented that the unrealistic or artificial content may bring confusion to the audience.
We further discuss the future research induced by this finding in ~\autoref{sec:discussion}

\subsubsection{Interacting with Gen-AI is easy; Guiding it is hard}
When asked about the interactions with the testbed, many appreciated that this testbed required less effort and content could be generated faster than the creation by humans.
This provides a unique edge over existing platforms (P5, P23, P24, P29).
\textit{"I feel like even though I have I have to do multiple interactions but still the content we can get quickly. I have only seen searching as this quick."}
Also, co-creation with Gen-AI is fun, making participants more indulged. 
However, some reported that the generation of the outputs was not controllable, there were unexpected or unintended outputs, and some were ethically incorrect because of incorrect prompts.
Many participants find it difficult to generate the desired content because of incorrect and incomplete prompts.
\textit{"I had to perform many iterations to get the desired output. It would have been much better if there was a way my input could give the desired output in one or two iterations." (P5)} 

The workflow allows participants to explore various outputs in a short period.
This testbed generates more creative content (P23, P27).
\textit{I can't believe that I was able to get an extraordinary image of foggy nights, I feel like I found content generated far better than my imagination" (P23).} 
Participants found that the generated content was easily immersed in the AR environment and was supplementing it.
Most of the participants found that the generated content was aligned with their intentions and also with the story. 


\section{Discussion}
\label{sec:discussion}
In this section, we discuss the primary results of the study and contrast them with prior works.
Furthermore, we provide more insights into the future landscape of AIGC in AR Storytelling, as well as design considerations for future AR Storytelling Systems and Applications.

\subsection{Modalities of AIGC in AR Storytelling}
Our first research question was whether multi-modal AIGC is suitable for AR Storytelling.
In ~\autoref{sec:results}, we show results indicating users' preferences for modalities to augment their storytelling.
We showed how participants are positively satisfied with the overall quality of each modality in terms of expressiveness, alignment with their intention, immersion into telling and listening, and exploration of options.
From the results, we further identified open challenges to address in order to situate AIGC in AR Storytelling.
We discuss the challenges in the following three aspects.

\begin{figure}[htp]
    \centering
    \includegraphics[width=\linewidth]{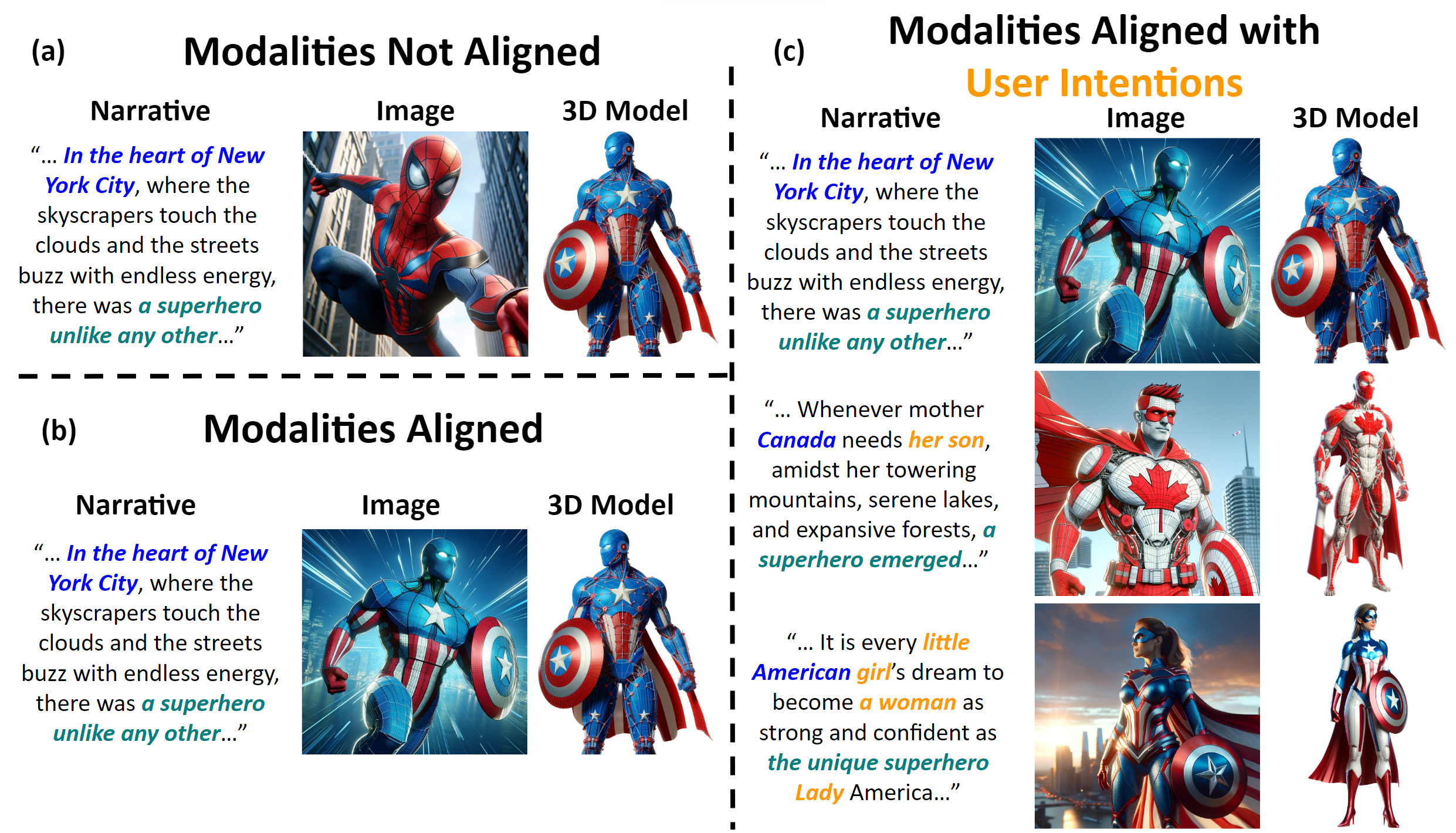}
    \caption{An illustration of the Alignment among modalities in AR Storytelling. All images are generated by ChatGPT~\protect\footnotemark with proper prompts. In (a), the image shows a different character from that shown by the 3D model, even though they are generated from the same textual narrative. In (b), we prompt the image generation with a screenshot of the 3D model, then the generated image is aligned with consistent character information from the 3D model. In (c), we prompted the image and 3D generation with user-intended information, such as nationality and gender, in order to obtain AIGC aligned among modalities as well as the users.}
    \label{fig:discussion-1}
\end{figure}
\footnotetext{\url{https://chat.openai.com/}}
\subsubsection{Alignment Among the Modalities and the Users' Intention}
Although single augmentation for a single element meets the satisfaction of the participants, it has been reported that the multi-modal augmentations for a single element appeared unaligned, i.e. the information of the element conveyed in one modality is different from that conveyed in another.
``\textit{I used both the image and the 3D model to introduce the character of a painter. I thought it would be cool to have the static image on my left hand and the rotating model of it on my right hand, but the model showed a different character from the image. (P21)}''
This finding echos with one heated field of Gen-AI research, where prior works~\cite{bommasani2021opportunities, devlin2018bert, brown2020language, radford2021learning} propose to train models on a large scale of data to be adapted into various downstream tasks.
By sharing the latent features and knowledge of the elements, the model can generate consistent multi-modal augmentations of the same elements from the story.
Yet, to bridge between the Gen-AI research and the realm of AR Storytelling, we anticipate the future one step further, where the human intention is also aligned within the co-creation process, by \textbf{\textit{evaluating not only the alignment among the modalities but also the modalities and the users' intention}}~\cite{antony2023id, lawton2023drawing}.
The alignment with the users is rather important in addressing the identity (e.g., race, gender, youth, or nationality) of both the tellers and the audience~\cite{brekke2021address, rolon2011race, bell2010storytelling}.


\begin{figure}[htp]
    \centering
    \includegraphics[width=\linewidth]{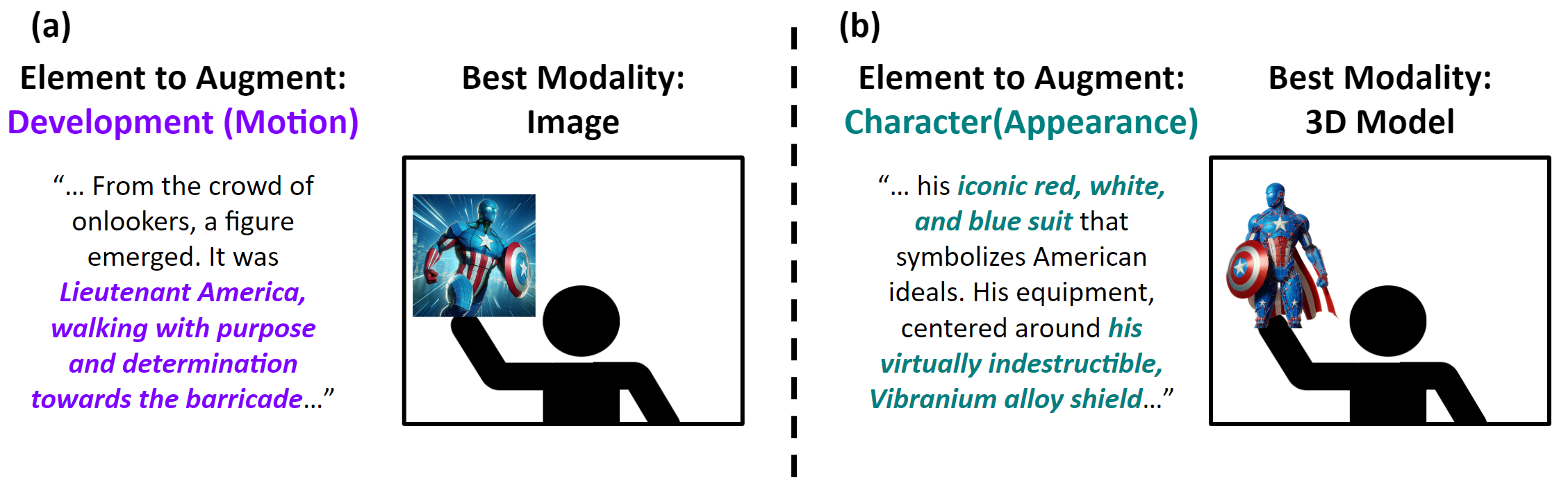}
    \caption{An illustration of the concept of Selective Augmentation. In (a), the narrative depicts the motion of a character entering the scene. In (b), the narrative introduces the appearance of a character. Selective Augmentation suggests that the system should infer the best modality to augment this narrative, without excessively augmenting it with too many modalities or poorly augmenting it with a less supportive modality.}
    \label{fig:discussion-2}
\end{figure}

\subsubsection{Selective Augmentation - What to Augment}
Despite the consistently positive ratings of AIGC augmentations, we have witnessed differences between the choices of elements in Study 2 (freestyle) and the elements we have assigned in Study 1.
In Study 2 where the choices of elements to augment were free, participants tended to augment fewer elements than we assigned in Study 1.
``\textit{Not all elements are necessarily needed to be augmented. Also, I do not tend to use multiple modalities to augment one simple element. For example for the background, just an image will do. I do not want too many visuals to show up in my presentation and confuse my audience. (P27)}''
This finding aligns with the fact that (1) storytellers possess individual preferences and strategies for choosing what to emphasize in their stories~\cite{walwema2015art}, that (2) excessive augmentation leads to convoluted presentation of the story~\cite{liao2022realitytalk}, and that (3) redundant information adds up to the mental cost of the audience to understand the story~\cite{peterson2006communication}.
Prior works~\cite{antony2023id, bensaid2021fairytailor} have suggested tackling the problem by adapting to individual co-creation in visual story generation, such that the system interprets what the authors would like to augment with visuals.

In addition, this adaptation should also consider the modality of the augmentation.
Given a system adaptive to and aware of what to augment, the expression is then constrained by the attributes of the element.
For example, in our study, we discovered that videos are preferred for the element of development, since ``\textit{...it (video) tells what the action is. No other option there could properly describe ``spent days and nights''. (P4)}''.
We also conclude from our research that the mapping of the best modalities to augment an element is not bijective.
Although video can show dynamic elements in the story, an image can sometimes be precise and vivid enough to depict a monotonous motion and is thus preferred over videos.
It is significantly important to understand the purpose of the elements in the story, in order to select the best modality or modalities.
These above findings inspire us to design authoring systems for AR Storytelling that adaptively infer \textbf{\textit{what to augment}} (in what modalities) based on user input and the narratives of the story.


\begin{figure}[htp]
    \centering
    \includegraphics[width=0.7\linewidth]{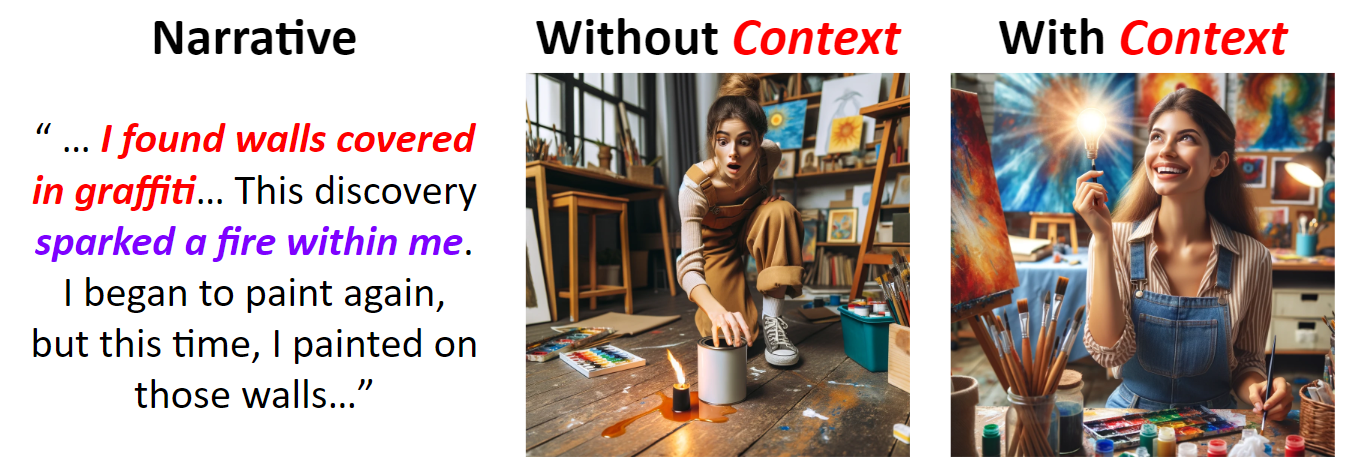}
    \caption{An illustration of context-awareness AIGC in AR Storytelling. Given a narrative, Gen-AI can generate illustrative images for storytelling. Prompting without the context of ``sparked a fire'' results in an image falsely illustrating the literal meaning, while prompting with the context of discovery artwork ``sparked a fire'' gives a correct illustration of the metaphor of an Eureka moment.}
    \label{fig:discussion-3}
\end{figure}

\subsubsection{Context-aware AIGC in AR Storytelling}
Among the four attributes we evaluated (expression, alignment, exploration, and immersion), we identified the lowest rating of the alignment between the AIGC and the participants' intention, which echos with the lowest Likert-scale score when the participants were asked about the easiness of prompting to guide the result generation.
These defects reflect that, on the one hand, the Gen-AI does not understand the context as the participants, ``\textit{... I would like to augment ``sparked a fire within me'' with an image depicting a moment of Eureka or something, but the image was generated with literally a fire being sparked inside... I don't think it understood metaphors. (P18)}''
Implicit expressions and metaphorical elements are proven to be engaging and leave room for audience imagination~\cite{taylor2018digital}.
Although prior works~\cite{wachowiak2023does} have demonstrated the ability of Gen-AI to comprehend metaphors, we argue that \textbf{\textit{context-awareness of the Gen-AI is needed for augmenting storytelling}}, enabling the Gen-AI to keep track of the elements of the story and understand what happens before and after the prompt.


\subsection{Interactions to Author AIGC in AR Storytelling}
Our second research question is to explore the key aspect when designing the interactions to create AIGC for AR Storytelling.
We showed in ~\autoref{sec:results} that co-creating with Gen-AI to augment Storytelling is easy in some modalities but faced with difficulty with some other modalities individually, and participants were overall satisfied with the ease and richness in variation compared to human-created content empirically.
Yet, our results showed that participants were not overwhelmingly positive about interactions during co-creation with Gen-AI, especially about how to guide the Gen-AI to generate their intended results.
From our analysis, we further discuss the following factors to be taken into consideration when designing the interactions.

\subsubsection{Interacting in AR to Prompt}
According to our results, participants gave the lowest rating when evaluating the easiness of prompting to guide the generation of the system.
In our experiment setup, we only enabled text-to-other content generation, where participants prompted with textual narratives.
For modalities that require more details, prompting with texts is not intuitive in guiding the generation.
``\textit{I had trouble generating the motion of someone wandering and then stumbling. I think the video is the best way to express this. I tried different text inputs of the same meaning, but none of the generated videos did it the way I pictured in my mind... I would rather have demonstrated it myself. (P2)}''
Prior works have demonstrated the potential applications of utilizing spatial interactions in AR to interact with general AI, e.g. authoring through hand gesture~\cite{wang2021gesturar}, sketch-to-image generation~\cite{bhattacharjee2020survey, ji2020sketch, duan2024conceptvis}, motion-to-animation~\cite{ye2020aranimator,shi2025caring}.
AR is a platform enabling multi-modalities not only for outputs but also inputs.
Inspired by the results, we anticipate future research on AIGC in AR to reveal more possibilities for \textbf{\textit{AR interactions prompting GenAI}} to enable intuitive content creation in AR, more than just text.

\subsubsection{Live Authoring During Storytelling - When to Augment}
In our study setup, we recruited experienced storytellers and presenters to augment their storytelling experience and examine it with our testbed.
We have witnessed some negative ratings of the experience's expressiveness and immersion.
As some participants pointed out, the audience is an essential part of the storytelling and their reaction is a deciding factor for the storytellers on when they should emphasize or elaborate part of the story.
``\textit{I think a better way to use this system is that when I actually tell the story and I figure out, from the audience's faces or my own improvisation, that I need something to show them so that they know what I am saying. You know, it is more of a feeling than some program. (P29)}''
Storytelling is a two-way communication~\cite{peterson2006communication}, even though the audience is silent most of the time.
It is shown by the prior works such as Visual Captions~\cite{liu2023visual} that when the content creation to augment the conversation is automated, communicators prefer live augmentations of their expression.
With Gen-AI taking the burden of content creation, we suggest that the storytellers have more freedom to decide \textbf{\textit{When to Augment}} their story during live presentations.
Further, we anticipate future systems to have the ability to actively infer and augment AR Storytelling when the presenters need it during live presentations.

\section{Study Limitations}
\label{sec:limitations}



For our study, we have developed a testbed consisting of a multi-modality content generator and an interactive AR Storytelling platform.
The collected study results are heavily based on users' experiences within our specific testbed environment.
Although the testbed serves the general purpose of exploring the effect of multi-modal AIGC in AR Storytelling, we did not design it in consideration of all specific or dedicated types of Storytelling tasks in diverse AR platforms.
We acknowledge that for a target task in AR Storytelling, there can be specific design considerations or requirements for the generated content, and thus dedicated metrics for evaluating task-related output.

Additionally, Our testbed is built upon a desktop application with a webcam.
The advances in AR headsets have enabled many interactions and experiences that are different from a desktop AR experience.
This difference may limit the ability to fully immerse users in a three-dimensional, interactive environment.
AR headsets can offer a hands-free experience.
In contrast, using a laptop camera might require users to stay in a specific position, limiting their mobility and potentially affecting the overall engagement during Storytelling in AR.
Most of the available multi-modal generative AI for AR Storytelling remains on desktop applications (e.g.~\cite{antony2023id, liu2023visual, chung2022talebrush}).
In this paper, we aim to answer \textbf{\textit{whether and why}} we should deploy multi-modal AIGC in AR Storytelling while acknowledging \textbf{\textit{how to do it}} on the most popular platforms as another open research question, which is ongoing in the field (e.g.~\cite{hu2023exploring, jonesusing, zhao2018compensation}) and also discussed in this paper. 
Yet, the contribution of our setup is a testbed for the study rather than a comprehensive system.
Although the choice of platform may change the spatial experience, we anticipate that the impact of AIGC on AR Storytelling can also be revealed via experiments on desktop-based AR\cite{madeira2022comparing, camba2014desktop}.

Lastly, we focus on the modalities of the content in AR Storytelling, while the interactivity of AR Storytelling is also a key and promising aspect that we do not comprehensively cover.
Such interactions involve (1) the necessary interactions the storytellers need to generate the content (during generation) as well as (2) the interactions between the tellers and the authored content to understand the story (during presentation).
With a focus on the modalities of AIGC, we mainly investigated (1) the \textbf{\textit{interactions to create the content}}.
It is essential to recognize that Storytelling can manifest in diverse forms, and the limitations associated with hand interaction may influence the observed limitation.
Thus, we suggest interpreting the results as a reference for utilizing Gen-AI in the creation of multi-modality content specifically for AR Storytelling with hand interaction.
We anticipate future research to explore the applicability of these findings to other interaction modalities in AR Storytelling or to broader contexts beyond the scope of our developed testbed.


\section{Conclusion}
In this paper, we have presented an exploratory study of Mutli-modal Gen-AI in AR Storytelling. 
We created an AR-based test bed that is capable of creating multi-modal content as well as using the content in AR for storytelling. 
We initially reviewed 223 videos to find the design space for AR storytelling. 
We found two dimensions: 1) Modalities - text, audio, image, video, and 3D and 2) Elements - Character, Background, Sentiment, and Development. 
We conducted two studies to evaluate the effect of multi-modal generated content in AR Storytelling.
We invited 30 participants, who were then divided into 2 groups of 15 each.
From the analysis of both qualitative and quantitative results of the study, we provide insights into the future directions of AIGC in AR storytelling.
We further discuss design considerations for future AR storytelling systems.

\bibliographystyle{ACM-Reference-Format}
\bibliography{sample-base}


\begin{thebibliography}{115}


\ifx \showCODEN    \undefined \def \showCODEN     #1{\unskip}     \fi
\ifx \showDOI      \undefined \def \showDOI       #1{#1}\fi
\ifx \showISBNx    \undefined \def \showISBNx     #1{\unskip}     \fi
\ifx \showISBNxiii \undefined \def \showISBNxiii  #1{\unskip}     \fi
\ifx \showISSN     \undefined \def \showISSN      #1{\unskip}     \fi
\ifx \showLCCN     \undefined \def \showLCCN      #1{\unskip}     \fi
\ifx \shownote     \undefined \def \shownote      #1{#1}          \fi
\ifx \showarticletitle \undefined \def \showarticletitle #1{#1}   \fi
\ifx \showURL      \undefined \def \showURL       {\relax}        \fi
\providecommand\bibfield[2]{#2}
\providecommand\bibinfo[2]{#2}
\providecommand\natexlab[1]{#1}
\providecommand\showeprint[2][]{arXiv:#2}

\bibitem[Abdal et~al\mbox{.}(2022)]%
        {abdal2022clip2stylegan}
\bibfield{author}{\bibinfo{person}{Rameen Abdal}, \bibinfo{person}{Peihao Zhu}, \bibinfo{person}{John Femiani}, \bibinfo{person}{Niloy Mitra}, {and} \bibinfo{person}{Peter Wonka}.} \bibinfo{year}{2022}\natexlab{}.
\newblock \showarticletitle{Clip2stylegan: Unsupervised extraction of stylegan edit directions}. In \bibinfo{booktitle}{\emph{ACM SIGGRAPH 2022 conference proceedings}}. \bibinfo{pages}{1--9}.
\newblock


\bibitem[Ahuja and Morency(2019)]%
        {ahuja2019language2pose}
\bibfield{author}{\bibinfo{person}{Chaitanya Ahuja} {and} \bibinfo{person}{Louis-Philippe Morency}.} \bibinfo{year}{2019}\natexlab{}.
\newblock \showarticletitle{Language2pose: Natural language grounded pose forecasting}. In \bibinfo{booktitle}{\emph{2019 International Conference on 3D Vision (3DV)}}. IEEE, \bibinfo{pages}{719--728}.
\newblock


\bibitem[Antony and Huang(2023)]%
        {antony2023id}
\bibfield{author}{\bibinfo{person}{Victor~Nikhil Antony} {and} \bibinfo{person}{Chien-Ming Huang}.} \bibinfo{year}{2023}\natexlab{}.
\newblock \showarticletitle{ID. 8: Co-Creating Visual Stories with Generative AI}.
\newblock \bibinfo{journal}{\emph{arXiv preprint arXiv:2309.14228}} (\bibinfo{year}{2023}).
\newblock


\bibitem[Association et~al\mbox{.}(2012)]%
        {national2012storytelling}
\bibfield{author}{\bibinfo{person}{National~Storytelling Association} {et~al\mbox{.}}} \bibinfo{year}{2012}\natexlab{}.
\newblock \bibinfo{title}{What Storytelling is. An attempt at defining the art form}.
\newblock
\newblock


\bibitem[Bala et~al\mbox{.}(2022)]%
        {bala2022writing}
\bibfield{author}{\bibinfo{person}{Paulo Bala}, \bibinfo{person}{Stuart James}, \bibinfo{person}{Alessio Del~Bue}, {and} \bibinfo{person}{Valentina Nisi}.} \bibinfo{year}{2022}\natexlab{}.
\newblock \showarticletitle{Writing with (Digital) Scissors: Designing a Text Editing Tool for Assisted Storytelling Using Crowd-Generated Content}. In \bibinfo{booktitle}{\emph{International Conference on Interactive Digital Storytelling}}. Springer, \bibinfo{pages}{139--158}.
\newblock


\bibitem[Bauer et~al\mbox{.}(2019)]%
        {bauer2019designing}
\bibfield{author}{\bibinfo{person}{Valentin Bauer}, \bibinfo{person}{Anna Nagele}, \bibinfo{person}{Chris Baume}, \bibinfo{person}{Tim Cowlishaw}, \bibinfo{person}{Henry Cooke}, \bibinfo{person}{Chris Pike}, {and} \bibinfo{person}{Patrick~GT Healey}.} \bibinfo{year}{2019}\natexlab{}.
\newblock \showarticletitle{Designing an interactive and collaborative experience in audio augmented reality}. In \bibinfo{booktitle}{\emph{Virtual Reality and Augmented Reality: 16th EuroVR International Conference, EuroVR 2019, Tallinn, Estonia, October 23--25, 2019, Proceedings 16}}. Springer, \bibinfo{pages}{305--311}.
\newblock


\bibitem[Bell(2010)]%
        {bell2010storytelling}
\bibfield{author}{\bibinfo{person}{Lee~Anne Bell}.} \bibinfo{year}{2010}\natexlab{}.
\newblock \bibinfo{booktitle}{\emph{Storytelling for social justice: Connecting narrative and the arts in antiracist teaching}}.
\newblock \bibinfo{publisher}{Routledge}.
\newblock


\bibitem[Bensaid et~al\mbox{.}(2021)]%
        {bensaid2021fairytailor}
\bibfield{author}{\bibinfo{person}{Eden Bensaid}, \bibinfo{person}{Mauro Martino}, \bibinfo{person}{Benjamin Hoover}, {and} \bibinfo{person}{Hendrik Strobelt}.} \bibinfo{year}{2021}\natexlab{}.
\newblock \showarticletitle{Fairytailor: A multimodal generative framework for storytelling}.
\newblock \bibinfo{journal}{\emph{arXiv preprint arXiv:2108.04324}} (\bibinfo{year}{2021}).
\newblock


\bibitem[Bhattacharjee and Chaudhuri(2020)]%
        {bhattacharjee2020survey}
\bibfield{author}{\bibinfo{person}{Sukanya Bhattacharjee} {and} \bibinfo{person}{Parag Chaudhuri}.} \bibinfo{year}{2020}\natexlab{}.
\newblock \showarticletitle{A survey on sketch based content creation: from the desktop to virtual and augmented reality}. In \bibinfo{booktitle}{\emph{Computer Graphics Forum}}, Vol.~\bibinfo{volume}{39}. Wiley Online Library, \bibinfo{pages}{757--780}.
\newblock


\bibitem[Bi{\'n}kowski et~al\mbox{.}(2019)]%
        {binkowski2019high}
\bibfield{author}{\bibinfo{person}{Miko{\l}aj Bi{\'n}kowski}, \bibinfo{person}{Jeff Donahue}, \bibinfo{person}{Sander Dieleman}, \bibinfo{person}{Aidan Clark}, \bibinfo{person}{Erich Elsen}, \bibinfo{person}{Norman Casagrande}, \bibinfo{person}{Luis~C Cobo}, {and} \bibinfo{person}{Karen Simonyan}.} \bibinfo{year}{2019}\natexlab{}.
\newblock \showarticletitle{High fidelity speech synthesis with adversarial networks}.
\newblock \bibinfo{journal}{\emph{arXiv preprint arXiv:1909.11646}} (\bibinfo{year}{2019}).
\newblock


\bibitem[Blattmann et~al\mbox{.}(2023)]%
        {blattmann2023align}
\bibfield{author}{\bibinfo{person}{Andreas Blattmann}, \bibinfo{person}{Robin Rombach}, \bibinfo{person}{Huan Ling}, \bibinfo{person}{Tim Dockhorn}, \bibinfo{person}{Seung~Wook Kim}, \bibinfo{person}{Sanja Fidler}, {and} \bibinfo{person}{Karsten Kreis}.} \bibinfo{year}{2023}\natexlab{}.
\newblock \showarticletitle{Align your latents: High-resolution video synthesis with latent diffusion models}. In \bibinfo{booktitle}{\emph{Proceedings of the IEEE/CVF Conference on Computer Vision and Pattern Recognition}}. \bibinfo{pages}{22563--22575}.
\newblock


\bibitem[Bommasani et~al\mbox{.}(2021)]%
        {bommasani2021opportunities}
\bibfield{author}{\bibinfo{person}{Rishi Bommasani}, \bibinfo{person}{Drew~A Hudson}, \bibinfo{person}{Ehsan Adeli}, \bibinfo{person}{Russ Altman}, \bibinfo{person}{Simran Arora}, \bibinfo{person}{Sydney von Arx}, \bibinfo{person}{Michael~S Bernstein}, \bibinfo{person}{Jeannette Bohg}, \bibinfo{person}{Antoine Bosselut}, \bibinfo{person}{Emma Brunskill}, {et~al\mbox{.}}} \bibinfo{year}{2021}\natexlab{}.
\newblock \showarticletitle{On the opportunities and risks of foundation models}.
\newblock \bibinfo{journal}{\emph{arXiv preprint arXiv:2108.07258}} (\bibinfo{year}{2021}).
\newblock


\bibitem[Brekke et~al\mbox{.}(2021)]%
        {brekke2021address}
\bibfield{author}{\bibinfo{person}{Anjuli~Joshi Brekke}, \bibinfo{person}{Ralina Joseph}, {and} \bibinfo{person}{Naheed~Gina Aaftaab}.} \bibinfo{year}{2021}\natexlab{}.
\newblock \showarticletitle{“I address race because race addresses me”: women of color show receipts through digital storytelling}.
\newblock \bibinfo{journal}{\emph{Review of Communication}} \bibinfo{volume}{21}, \bibinfo{number}{1} (\bibinfo{year}{2021}), \bibinfo{pages}{44--57}.
\newblock


\bibitem[Brown et~al\mbox{.}(2020)]%
        {brown2020language}
\bibfield{author}{\bibinfo{person}{Tom Brown}, \bibinfo{person}{Benjamin Mann}, \bibinfo{person}{Nick Ryder}, \bibinfo{person}{Melanie Subbiah}, \bibinfo{person}{Jared~D Kaplan}, \bibinfo{person}{Prafulla Dhariwal}, \bibinfo{person}{Arvind Neelakantan}, \bibinfo{person}{Pranav Shyam}, \bibinfo{person}{Girish Sastry}, \bibinfo{person}{Amanda Askell}, {et~al\mbox{.}}} \bibinfo{year}{2020}\natexlab{}.
\newblock \showarticletitle{Language models are few-shot learners}.
\newblock \bibinfo{journal}{\emph{Advances in neural information processing systems}}  \bibinfo{volume}{33} (\bibinfo{year}{2020}), \bibinfo{pages}{1877--1901}.
\newblock


\bibitem[Calvi(2020)]%
        {calvi2020we}
\bibfield{author}{\bibinfo{person}{Licia Calvi}.} \bibinfo{year}{2020}\natexlab{}.
\newblock \showarticletitle{What do we know about AR Storytelling?}. In \bibinfo{booktitle}{\emph{Proceedings of the 6th EAI International Conference on Smart Objects and Technologies for Social Good}}. \bibinfo{pages}{278--280}.
\newblock


\bibitem[Camba et~al\mbox{.}(2014)]%
        {camba2014desktop}
\bibfield{author}{\bibinfo{person}{Jorge Camba}, \bibinfo{person}{Manuel Contero}, {and} \bibinfo{person}{Gustavo Salvador-Herranz}.} \bibinfo{year}{2014}\natexlab{}.
\newblock \showarticletitle{Desktop vs. mobile: A comparative study of augmented reality systems for engineering visualizations in education}. In \bibinfo{booktitle}{\emph{2014 IEEE Frontiers in Education Conference (FIE) Proceedings}}. IEEE, \bibinfo{pages}{1--8}.
\newblock


\bibitem[Chen et~al\mbox{.}(2023)]%
        {chen2023control}
\bibfield{author}{\bibinfo{person}{Weifeng Chen}, \bibinfo{person}{Jie Wu}, \bibinfo{person}{Pan Xie}, \bibinfo{person}{Hefeng Wu}, \bibinfo{person}{Jiashi Li}, \bibinfo{person}{Xin Xia}, \bibinfo{person}{Xuefeng Xiao}, {and} \bibinfo{person}{Liang Lin}.} \bibinfo{year}{2023}\natexlab{}.
\newblock \showarticletitle{Control-A-Video: Controllable Text-to-Video Generation with Diffusion Models}.
\newblock \bibinfo{journal}{\emph{arXiv preprint arXiv:2305.13840}} (\bibinfo{year}{2023}).
\newblock


\bibitem[Chung et~al\mbox{.}(2022)]%
        {chung2022talebrush}
\bibfield{author}{\bibinfo{person}{John Joon~Young Chung}, \bibinfo{person}{Wooseok Kim}, \bibinfo{person}{Kang~Min Yoo}, \bibinfo{person}{Hwaran Lee}, \bibinfo{person}{Eytan Adar}, {and} \bibinfo{person}{Minsuk Chang}.} \bibinfo{year}{2022}\natexlab{}.
\newblock \showarticletitle{TaleBrush: visual sketching of story generation with pretrained language models}. In \bibinfo{booktitle}{\emph{CHI Conference on Human Factors in Computing Systems Extended Abstracts}}. \bibinfo{pages}{1--4}.
\newblock


\bibitem[Copet et~al\mbox{.}(2023)]%
        {copet2023simple}
\bibfield{author}{\bibinfo{person}{Jade Copet}, \bibinfo{person}{Felix Kreuk}, \bibinfo{person}{Itai Gat}, \bibinfo{person}{Tal Remez}, \bibinfo{person}{David Kant}, \bibinfo{person}{Gabriel Synnaeve}, \bibinfo{person}{Yossi Adi}, {and} \bibinfo{person}{Alexandre D{\'e}fossez}.} \bibinfo{year}{2023}\natexlab{}.
\newblock \showarticletitle{Simple and Controllable Music Generation}.
\newblock \bibinfo{journal}{\emph{arXiv preprint arXiv:2306.05284}} (\bibinfo{year}{2023}).
\newblock


\bibitem[Danaei et~al\mbox{.}(2020)]%
        {danaei2020comparing}
\bibfield{author}{\bibinfo{person}{Delneshin Danaei}, \bibinfo{person}{Hamid~R Jamali}, \bibinfo{person}{Yazdan Mansourian}, {and} \bibinfo{person}{Hassan Rastegarpour}.} \bibinfo{year}{2020}\natexlab{}.
\newblock \showarticletitle{Comparing reading comprehension between children reading augmented reality and print storybooks}.
\newblock \bibinfo{journal}{\emph{Computers \& Education}}  \bibinfo{volume}{153} (\bibinfo{year}{2020}), \bibinfo{pages}{103900}.
\newblock


\bibitem[Darzentas et~al\mbox{.}(2018)]%
        {darzentas2018object}
\bibfield{author}{\bibinfo{person}{Dimitrios Darzentas}, \bibinfo{person}{Martin Flintham}, {and} \bibinfo{person}{Steve Benford}.} \bibinfo{year}{2018}\natexlab{}.
\newblock \showarticletitle{Object-focused mixed reality storytelling: technology-driven content creation and dissemination for engaging user experiences}. In \bibinfo{booktitle}{\emph{Proceedings of the 22nd Pan-Hellenic Conference on Informatics}}. \bibinfo{pages}{278--281}.
\newblock


\bibitem[Devlin et~al\mbox{.}(2018)]%
        {devlin2018bert}
\bibfield{author}{\bibinfo{person}{Jacob Devlin}, \bibinfo{person}{Ming-Wei Chang}, \bibinfo{person}{Kenton Lee}, {and} \bibinfo{person}{Kristina Toutanova}.} \bibinfo{year}{2018}\natexlab{}.
\newblock \showarticletitle{Bert: Pre-training of deep bidirectional transformers for language understanding}.
\newblock \bibinfo{journal}{\emph{arXiv preprint arXiv:1810.04805}} (\bibinfo{year}{2018}).
\newblock


\bibitem[Ding et~al\mbox{.}(2022)]%
        {ding2022cogview2}
\bibfield{author}{\bibinfo{person}{Ming Ding}, \bibinfo{person}{Wendi Zheng}, \bibinfo{person}{Wenyi Hong}, {and} \bibinfo{person}{Jie Tang}.} \bibinfo{year}{2022}\natexlab{}.
\newblock \showarticletitle{Cogview2: Faster and better text-to-image generation via hierarchical transformers}.
\newblock \bibinfo{journal}{\emph{Advances in Neural Information Processing Systems}}  \bibinfo{volume}{35} (\bibinfo{year}{2022}), \bibinfo{pages}{16890--16902}.
\newblock


\bibitem[Duan et~al\mbox{.}(2025a)]%
        {duan2025investigating}
\bibfield{author}{\bibinfo{person}{Runlin Duan}, \bibinfo{person}{Shao-Kang Hsia}, \bibinfo{person}{Yuzhao Chen}, \bibinfo{person}{Yichen Hu}, \bibinfo{person}{Ming Yin}, {and} \bibinfo{person}{Karthik Ramani}.} \bibinfo{year}{2025}\natexlab{a}.
\newblock \showarticletitle{Investigating Creativity in Humans and Generative AI Through Circles Exercises}.
\newblock \bibinfo{journal}{\emph{arXiv preprint arXiv:2502.07292}} (\bibinfo{year}{2025}).
\newblock


\bibitem[Duan et~al\mbox{.}(2024)]%
        {duan2024conceptvis}
\bibfield{author}{\bibinfo{person}{Runlin Duan}, \bibinfo{person}{Nachiketh Karthik}, \bibinfo{person}{Jingyu Shi}, \bibinfo{person}{Rahul Jain}, \bibinfo{person}{Maria~C Yang}, {and} \bibinfo{person}{Karthik Ramani}.} \bibinfo{year}{2024}\natexlab{}.
\newblock \showarticletitle{ConceptVis: Generating and Exploring Design Concepts for Early-Stage Ideation Using Large Language Model}. In \bibinfo{booktitle}{\emph{International Design Engineering Technical Conferences and Computers and Information in Engineering Conference}}, Vol.~\bibinfo{volume}{88377}. American Society of Mechanical Engineers, \bibinfo{pages}{V03BT03A042}.
\newblock


\bibitem[Duan et~al\mbox{.}(2025b)]%
        {duan2025designfromx}
\bibfield{author}{\bibinfo{person}{Runlin Duan}, \bibinfo{person}{Chenfei Zhu}, \bibinfo{person}{Yuzhao Chen}, \bibinfo{person}{Yichen Hu}, \bibinfo{person}{Jingyu Shi}, {and} \bibinfo{person}{Karthik Ramani}.} \bibinfo{year}{2025}\natexlab{b}.
\newblock \showarticletitle{DesignFromX: Empowering Consumer-Driven Design Space Exploration through Feature Composition of Referenced Products}.
\newblock \bibinfo{journal}{\emph{arXiv preprint arXiv:2505.11666}} (\bibinfo{year}{2025}).
\newblock


\bibitem[FastAPI({[n.\,d.]})]%
        {fastapi}
\bibfield{author}{\bibinfo{person}{FastAPI}.} \bibinfo{year}{[n.\,d.]}\natexlab{}.
\newblock \bibinfo{title}{FastAPI}.
\newblock \bibinfo{howpublished}{\url{https://fastapi.tiangolo.com}}.
\newblock


\bibitem[Fog et~al\mbox{.}(2005)]%
        {fog2005storytelling}
\bibfield{author}{\bibinfo{person}{Klaus Fog}, \bibinfo{person}{Christian Budtz}, {and} \bibinfo{person}{Baris Yakaboylu}.} \bibinfo{year}{2005}\natexlab{}.
\newblock \bibinfo{booktitle}{\emph{Storytelling}}.
\newblock \bibinfo{publisher}{Springer}.
\newblock


\bibitem[Gal et~al\mbox{.}(2022)]%
        {gal2022stylegan}
\bibfield{author}{\bibinfo{person}{Rinon Gal}, \bibinfo{person}{Or Patashnik}, \bibinfo{person}{Haggai Maron}, \bibinfo{person}{Amit~H Bermano}, \bibinfo{person}{Gal Chechik}, {and} \bibinfo{person}{Daniel Cohen-Or}.} \bibinfo{year}{2022}\natexlab{}.
\newblock \showarticletitle{StyleGAN-NADA: CLIP-guided domain adaptation of image generators}.
\newblock \bibinfo{journal}{\emph{ACM Transactions on Graphics (TOG)}} \bibinfo{volume}{41}, \bibinfo{number}{4} (\bibinfo{year}{2022}), \bibinfo{pages}{1--13}.
\newblock


\bibitem[Gao et~al\mbox{.}(2022)]%
        {gao2022bridging}
\bibfield{author}{\bibinfo{person}{Ze Gao}, \bibinfo{person}{Anqi Wang}, \bibinfo{person}{Pan Hui}, {and} \bibinfo{person}{Tristan Braud}.} \bibinfo{year}{2022}\natexlab{}.
\newblock \showarticletitle{Bridging curatorial intent and visiting experience: Using ar guidance as a storytelling tool}. In \bibinfo{booktitle}{\emph{Proceedings of the 18th ACM SIGGRAPH International Conference on Virtual-Reality Continuum and its Applications in Industry}}. \bibinfo{pages}{1--10}.
\newblock


\bibitem[Goodfellow et~al\mbox{.}(2014)]%
        {goodfellow2014generative}
\bibfield{author}{\bibinfo{person}{Ian Goodfellow}, \bibinfo{person}{Jean Pouget-Abadie}, \bibinfo{person}{Mehdi Mirza}, \bibinfo{person}{Bing Xu}, \bibinfo{person}{David Warde-Farley}, \bibinfo{person}{Sherjil Ozair}, \bibinfo{person}{Aaron Courville}, {and} \bibinfo{person}{Yoshua Bengio}.} \bibinfo{year}{2014}\natexlab{}.
\newblock \showarticletitle{Generative adversarial nets}.
\newblock \bibinfo{journal}{\emph{Advances in neural information processing systems}}  \bibinfo{volume}{27} (\bibinfo{year}{2014}).
\newblock


\bibitem[Han and Cai(2023)]%
        {han2023design}
\bibfield{author}{\bibinfo{person}{Ariel Han} {and} \bibinfo{person}{Zhenyao Cai}.} \bibinfo{year}{2023}\natexlab{}.
\newblock \showarticletitle{Design implications of generative AI systems for visual storytelling for young learners}. In \bibinfo{booktitle}{\emph{Proceedings of the 22nd Annual ACM Interaction Design and Children Conference}}. \bibinfo{pages}{470--474}.
\newblock


\bibitem[Hirsch et~al\mbox{.}(2022)]%
        {hirsch2022embedded}
\bibfield{author}{\bibinfo{person}{Linda Hirsch}, \bibinfo{person}{Robin Welsch}, \bibinfo{person}{Beat Rossmy}, {and} \bibinfo{person}{Andreas Butz}.} \bibinfo{year}{2022}\natexlab{}.
\newblock \showarticletitle{Embedded AR Storytelling Supports Active Indexing at Historical Places}. In \bibinfo{booktitle}{\emph{Sixteenth International Conference on Tangible, Embedded, and Embodied Interaction}}. \bibinfo{pages}{1--12}.
\newblock


\bibitem[Ho et~al\mbox{.}(2022)]%
        {ho2022imagen}
\bibfield{author}{\bibinfo{person}{Jonathan Ho}, \bibinfo{person}{William Chan}, \bibinfo{person}{Chitwan Saharia}, \bibinfo{person}{Jay Whang}, \bibinfo{person}{Ruiqi Gao}, \bibinfo{person}{Alexey Gritsenko}, \bibinfo{person}{Diederik~P Kingma}, \bibinfo{person}{Ben Poole}, \bibinfo{person}{Mohammad Norouzi}, \bibinfo{person}{David~J Fleet}, {et~al\mbox{.}}} \bibinfo{year}{2022}\natexlab{}.
\newblock \showarticletitle{Imagen video: High definition video generation with diffusion models}.
\newblock \bibinfo{journal}{\emph{arXiv preprint arXiv:2210.02303}} (\bibinfo{year}{2022}).
\newblock


\bibitem[Holloway-Attaway and Vipsj{\"o}(2020)]%
        {holloway2020using}
\bibfield{author}{\bibinfo{person}{Lissa Holloway-Attaway} {and} \bibinfo{person}{Lars Vipsj{\"o}}.} \bibinfo{year}{2020}\natexlab{}.
\newblock \showarticletitle{Using augmented reality, gaming technologies, and transmedial storytelling to develop and co-design local cultural heritage experiences}.
\newblock \bibinfo{journal}{\emph{Visual Computing for Cultural Heritage}} (\bibinfo{year}{2020}), \bibinfo{pages}{177--204}.
\newblock


\bibitem[Honnibal et~al\mbox{.}(2020)]%
        {honnibal2020spacy}
\bibfield{author}{\bibinfo{person}{Matthew Honnibal}, \bibinfo{person}{Ines Montani}, \bibinfo{person}{Sofie Van~Landeghem}, \bibinfo{person}{Adriane Boyd}, {et~al\mbox{.}}} \bibinfo{year}{2020}\natexlab{}.
\newblock \showarticletitle{spaCy: Industrial-strength natural language processing in python}.
\newblock  (\bibinfo{year}{2020}).
\newblock


\bibitem[Hu et~al\mbox{.}(2025)]%
        {hu2025gesprompt}
\bibfield{author}{\bibinfo{person}{Xiyun Hu}, \bibinfo{person}{Dizhi Ma}, \bibinfo{person}{Fengming He}, \bibinfo{person}{Zhengzhe Zhu}, \bibinfo{person}{Shao-Kang Hsia}, \bibinfo{person}{Chenfei Zhu}, \bibinfo{person}{Ziyi Liu}, {and} \bibinfo{person}{Karthik Ramani}.} \bibinfo{year}{2025}\natexlab{}.
\newblock \showarticletitle{GesPrompt: Leveraging Co-Speech Gestures to Augment LLM-Based Interaction in Virtual Reality}.
\newblock \bibinfo{journal}{\emph{arXiv preprint arXiv:2505.05441}} (\bibinfo{year}{2025}).
\newblock


\bibitem[Hu et~al\mbox{.}(2023)]%
        {hu2023exploring}
\bibfield{author}{\bibinfo{person}{Yongquan Hu}, \bibinfo{person}{Mingyue Yuan}, \bibinfo{person}{Kaiqi Xian}, \bibinfo{person}{Don~Samitha Elvitigala}, {and} \bibinfo{person}{Aaron Quigley}.} \bibinfo{year}{2023}\natexlab{}.
\newblock \showarticletitle{Exploring the Design Space of Employing AI-Generated Content for Augmented Reality Display}.
\newblock \bibinfo{journal}{\emph{arXiv preprint arXiv:2303.16593}} (\bibinfo{year}{2023}).
\newblock


\bibitem[Indans et~al\mbox{.}(2019)]%
        {indans2019towards}
\bibfield{author}{\bibinfo{person}{Reinis Indans}, \bibinfo{person}{Eva Hauthal}, {and} \bibinfo{person}{Dirk Burghardt}.} \bibinfo{year}{2019}\natexlab{}.
\newblock \showarticletitle{Towards an audio-locative mobile application for immersive storytelling}.
\newblock \bibinfo{journal}{\emph{KN-Journal of Cartography and Geographic Information}}  \bibinfo{volume}{69} (\bibinfo{year}{2019}), \bibinfo{pages}{41--50}.
\newblock


\bibitem[Ji and Chun(2020)]%
        {ji2020sketch}
\bibfield{author}{\bibinfo{person}{Myunggeun Ji} {and} \bibinfo{person}{Junchul Chun}.} \bibinfo{year}{2020}\natexlab{}.
\newblock \showarticletitle{A Sketch-based 3D Object Retrieval Approach for Augmented Reality Models Using Deep Learning.}
\newblock \bibinfo{journal}{\emph{Journal of Korean Society for Internet Information}} \bibinfo{volume}{21}, \bibinfo{number}{1} (\bibinfo{year}{2020}).
\newblock


\bibitem[JONES et~al\mbox{.}({[n.\,d.]})]%
        {jonesusing}
\bibfield{author}{\bibinfo{person}{BRENNAN JONES}, \bibinfo{person}{YAN XU}, \bibinfo{person}{MARY~ANNE HOOD}, \bibinfo{person}{MOHAMMAD~SHAHIDUL KADER}, {and} \bibinfo{person}{HAMID EGHBALZADEH}.} \bibinfo{year}{[n.\,d.]}\natexlab{}.
\newblock \showarticletitle{Using Generative AI to Produce Situated Action Recommendations in Augmented Reality for High-Level Goals}.
\newblock  (\bibinfo{year}{[n.\,d.]}).
\newblock


\bibitem[Jung et~al\mbox{.}(2021)]%
        {jung2021blocklyxr}
\bibfield{author}{\bibinfo{person}{Kwanghee Jung}, \bibinfo{person}{Vinh~T Nguyen}, {and} \bibinfo{person}{Jaehoon Lee}.} \bibinfo{year}{2021}\natexlab{}.
\newblock \showarticletitle{Blocklyxr: An interactive extended reality toolkit for digital storytelling}.
\newblock \bibinfo{journal}{\emph{Applied Sciences}} \bibinfo{volume}{11}, \bibinfo{number}{3} (\bibinfo{year}{2021}), \bibinfo{pages}{1073}.
\newblock


\bibitem[Kalchbrenner et~al\mbox{.}(2018)]%
        {kalchbrenner2018efficient}
\bibfield{author}{\bibinfo{person}{Nal Kalchbrenner}, \bibinfo{person}{Erich Elsen}, \bibinfo{person}{Karen Simonyan}, \bibinfo{person}{Seb Noury}, \bibinfo{person}{Norman Casagrande}, \bibinfo{person}{Edward Lockhart}, \bibinfo{person}{Florian Stimberg}, \bibinfo{person}{Aaron Oord}, \bibinfo{person}{Sander Dieleman}, {and} \bibinfo{person}{Koray Kavukcuoglu}.} \bibinfo{year}{2018}\natexlab{}.
\newblock \showarticletitle{Efficient neural audio synthesis}. In \bibinfo{booktitle}{\emph{International Conference on Machine Learning}}. PMLR, \bibinfo{pages}{2410--2419}.
\newblock


\bibitem[Karras et~al\mbox{.}(2021)]%
        {karras2021alias}
\bibfield{author}{\bibinfo{person}{Tero Karras}, \bibinfo{person}{Miika Aittala}, \bibinfo{person}{Samuli Laine}, \bibinfo{person}{Erik H{\"a}rk{\"o}nen}, \bibinfo{person}{Janne Hellsten}, \bibinfo{person}{Jaakko Lehtinen}, {and} \bibinfo{person}{Timo Aila}.} \bibinfo{year}{2021}\natexlab{}.
\newblock \showarticletitle{Alias-free generative adversarial networks}.
\newblock \bibinfo{journal}{\emph{Advances in Neural Information Processing Systems}}  \bibinfo{volume}{34} (\bibinfo{year}{2021}), \bibinfo{pages}{852--863}.
\newblock


\bibitem[Karras et~al\mbox{.}(2019)]%
        {karras2019style}
\bibfield{author}{\bibinfo{person}{Tero Karras}, \bibinfo{person}{Samuli Laine}, {and} \bibinfo{person}{Timo Aila}.} \bibinfo{year}{2019}\natexlab{}.
\newblock \showarticletitle{A style-based generator architecture for generative adversarial networks}. In \bibinfo{booktitle}{\emph{Proceedings of the IEEE/CVF conference on computer vision and pattern recognition}}. \bibinfo{pages}{4401--4410}.
\newblock


\bibitem[Ketchell et~al\mbox{.}(2019)]%
        {ketchell2019situated}
\bibfield{author}{\bibinfo{person}{Sarah Ketchell}, \bibinfo{person}{Winyu Chinthammit}, {and} \bibinfo{person}{Ulrich Engelke}.} \bibinfo{year}{2019}\natexlab{}.
\newblock \showarticletitle{Situated storytelling with SLAM enabled augmented reality}. In \bibinfo{booktitle}{\emph{Proceedings of the 17th International Conference on Virtual-Reality Continuum and Its Applications in Industry}}. \bibinfo{pages}{1--9}.
\newblock


\bibitem[Khachatryan et~al\mbox{.}(2023)]%
        {khachatryan2023text2video}
\bibfield{author}{\bibinfo{person}{Levon Khachatryan}, \bibinfo{person}{Andranik Movsisyan}, \bibinfo{person}{Vahram Tadevosyan}, \bibinfo{person}{Roberto Henschel}, \bibinfo{person}{Zhangyang Wang}, \bibinfo{person}{Shant Navasardyan}, {and} \bibinfo{person}{Humphrey Shi}.} \bibinfo{year}{2023}\natexlab{}.
\newblock \showarticletitle{Text2video-zero: Text-to-image diffusion models are zero-shot video generators}.
\newblock \bibinfo{journal}{\emph{arXiv preprint arXiv:2303.13439}} (\bibinfo{year}{2023}).
\newblock


\bibitem[Kim et~al\mbox{.}(2020)]%
        {kim2020tivgan}
\bibfield{author}{\bibinfo{person}{Doyeon Kim}, \bibinfo{person}{Donggyu Joo}, {and} \bibinfo{person}{Junmo Kim}.} \bibinfo{year}{2020}\natexlab{}.
\newblock \showarticletitle{Tivgan: Text to image to video generation with step-by-step evolutionary generator}.
\newblock \bibinfo{journal}{\emph{IEEE Access}}  \bibinfo{volume}{8} (\bibinfo{year}{2020}), \bibinfo{pages}{153113--153122}.
\newblock


\bibitem[Kim et~al\mbox{.}(2023)]%
        {kim2023flame}
\bibfield{author}{\bibinfo{person}{Jihoon Kim}, \bibinfo{person}{Jiseob Kim}, {and} \bibinfo{person}{Sungjoon Choi}.} \bibinfo{year}{2023}\natexlab{}.
\newblock \showarticletitle{Flame: Free-form language-based motion synthesis \& editing}. In \bibinfo{booktitle}{\emph{Proceedings of the AAAI Conference on Artificial Intelligence}}, Vol.~\bibinfo{volume}{37}. \bibinfo{pages}{8255--8263}.
\newblock


\bibitem[Kim et~al\mbox{.}(2014)]%
        {kim2014vision}
\bibfield{author}{\bibinfo{person}{Kiyoung Kim}, \bibinfo{person}{Noh-young Park}, {and} \bibinfo{person}{Woontack Woo}.} \bibinfo{year}{2014}\natexlab{}.
\newblock \showarticletitle{Vision-based all-in-one solution for augmented reality and its storytelling applications}.
\newblock \bibinfo{journal}{\emph{The Visual Computer}}  \bibinfo{volume}{30} (\bibinfo{year}{2014}), \bibinfo{pages}{417--429}.
\newblock


\bibitem[Kingma and Welling(2013)]%
        {kingma2013auto}
\bibfield{author}{\bibinfo{person}{Diederik~P Kingma} {and} \bibinfo{person}{Max Welling}.} \bibinfo{year}{2013}\natexlab{}.
\newblock \showarticletitle{Auto-encoding variational bayes}.
\newblock \bibinfo{journal}{\emph{arXiv preprint arXiv:1312.6114}} (\bibinfo{year}{2013}).
\newblock


\bibitem[Kong et~al\mbox{.}(2020)]%
        {kong2020diffwave}
\bibfield{author}{\bibinfo{person}{Zhifeng Kong}, \bibinfo{person}{Wei Ping}, \bibinfo{person}{Jiaji Huang}, \bibinfo{person}{Kexin Zhao}, {and} \bibinfo{person}{Bryan Catanzaro}.} \bibinfo{year}{2020}\natexlab{}.
\newblock \showarticletitle{Diffwave: A versatile diffusion model for audio synthesis}.
\newblock \bibinfo{journal}{\emph{arXiv preprint arXiv:2009.09761}} (\bibinfo{year}{2020}).
\newblock


\bibitem[Kumar et~al\mbox{.}(2019)]%
        {kumar2019melgan}
\bibfield{author}{\bibinfo{person}{Kundan Kumar}, \bibinfo{person}{Rithesh Kumar}, \bibinfo{person}{Thibault De~Boissiere}, \bibinfo{person}{Lucas Gestin}, \bibinfo{person}{Wei~Zhen Teoh}, \bibinfo{person}{Jose Sotelo}, \bibinfo{person}{Alexandre De~Brebisson}, \bibinfo{person}{Yoshua Bengio}, {and} \bibinfo{person}{Aaron~C Courville}.} \bibinfo{year}{2019}\natexlab{}.
\newblock \showarticletitle{Melgan: Generative adversarial networks for conditional waveform synthesis}.
\newblock \bibinfo{journal}{\emph{Advances in neural information processing systems}}  \bibinfo{volume}{32} (\bibinfo{year}{2019}).
\newblock


\bibitem[Lawton et~al\mbox{.}(2023)]%
        {lawton2023drawing}
\bibfield{author}{\bibinfo{person}{Tomas Lawton}, \bibinfo{person}{Francisco~J Ibarrola}, \bibinfo{person}{Dan Ventura}, {and} \bibinfo{person}{Kazjon Grace}.} \bibinfo{year}{2023}\natexlab{}.
\newblock \showarticletitle{Drawing with Reframer: Emergence and Control in Co-Creative AI}. In \bibinfo{booktitle}{\emph{Proceedings of the 28th International Conference on Intelligent User Interfaces}}. \bibinfo{pages}{264--277}.
\newblock


\bibitem[Lee et~al\mbox{.}(2023)]%
        {lee2023exploring}
\bibfield{author}{\bibinfo{person}{Kyungjun Lee}, \bibinfo{person}{Hong Li}, \bibinfo{person}{Muhammad~Rizky Wellyanto}, \bibinfo{person}{Yu~Jiang Tham}, \bibinfo{person}{Andr{\'e}s Monroy-Hern{\'a}ndez}, \bibinfo{person}{Fannie Liu}, \bibinfo{person}{Brian~A Smith}, {and} \bibinfo{person}{Rajan Vaish}.} \bibinfo{year}{2023}\natexlab{}.
\newblock \showarticletitle{Exploring Immersive Interpersonal Communication via AR}.
\newblock \bibinfo{journal}{\emph{Proceedings of the ACM on Human-Computer Interaction}} \bibinfo{volume}{7}, \bibinfo{number}{CSCW1} (\bibinfo{year}{2023}), \bibinfo{pages}{1--25}.
\newblock


\bibitem[Liao et~al\mbox{.}(2022)]%
        {liao2022realitytalk}
\bibfield{author}{\bibinfo{person}{Jian Liao}, \bibinfo{person}{Adnan Karim}, \bibinfo{person}{Shivesh~Singh Jadon}, \bibinfo{person}{Rubaiat~Habib Kazi}, {and} \bibinfo{person}{Ryo Suzuki}.} \bibinfo{year}{2022}\natexlab{}.
\newblock \showarticletitle{RealityTalk: Real-Time Speech-Driven Augmented Presentation for AR Live Storytelling}. In \bibinfo{booktitle}{\emph{Proceedings of the 35th Annual ACM Symposium on User Interface Software and Technology}}. \bibinfo{pages}{1--12}.
\newblock


\bibitem[Liu et~al\mbox{.}(2023)]%
        {liu2023visual}
\bibfield{author}{\bibinfo{person}{Xingyu"~Bruce" Liu}, \bibinfo{person}{Vladimir Kirilyuk}, \bibinfo{person}{Xiuxiu Yuan}, \bibinfo{person}{Alex Olwal}, \bibinfo{person}{Peggy Chi}, \bibinfo{person}{Xiang"~Anthony" Chen}, {and} \bibinfo{person}{Ruofei Du}.} \bibinfo{year}{2023}\natexlab{}.
\newblock \showarticletitle{Visual Captions: Augmenting Verbal Communication With On-the-Fly Visuals}. In \bibinfo{booktitle}{\emph{Proceedings of the 2023 CHI Conference on Human Factors in Computing Systems}}. \bibinfo{pages}{1--20}.
\newblock


\bibitem[Liu et~al\mbox{.}(2024)]%
        {liu2024classmeta}
\bibfield{author}{\bibinfo{person}{Ziyi Liu}, \bibinfo{person}{Zhengzhe Zhu}, \bibinfo{person}{Lijun Zhu}, \bibinfo{person}{Enze Jiang}, \bibinfo{person}{Xiyun Hu}, \bibinfo{person}{Kylie~A Peppler}, {and} \bibinfo{person}{Karthik Ramani}.} \bibinfo{year}{2024}\natexlab{}.
\newblock \showarticletitle{Classmeta: Designing interactive virtual classmate to promote VR classroom participation}. In \bibinfo{booktitle}{\emph{Proceedings of the 2024 CHI Conference on Human Factors in Computing Systems}}. \bibinfo{pages}{1--17}.
\newblock


\bibitem[Ma et~al\mbox{.}(2023)]%
        {ma2023follow}
\bibfield{author}{\bibinfo{person}{Yue Ma}, \bibinfo{person}{Yingqing He}, \bibinfo{person}{Xiaodong Cun}, \bibinfo{person}{Xintao Wang}, \bibinfo{person}{Ying Shan}, \bibinfo{person}{Xiu Li}, {and} \bibinfo{person}{Qifeng Chen}.} \bibinfo{year}{2023}\natexlab{}.
\newblock \showarticletitle{Follow Your Pose: Pose-Guided Text-to-Video Generation using Pose-Free Videos}.
\newblock \bibinfo{journal}{\emph{arXiv preprint arXiv:2304.01186}} (\bibinfo{year}{2023}).
\newblock


\bibitem[Madeira et~al\mbox{.}(2022)]%
        {madeira2022comparing}
\bibfield{author}{\bibinfo{person}{Tiago Madeira}, \bibinfo{person}{Bernardo Marques}, \bibinfo{person}{Pedro Neves}, \bibinfo{person}{Paulo Dias}, {and} \bibinfo{person}{Beatriz~Sousa Santos}.} \bibinfo{year}{2022}\natexlab{}.
\newblock \showarticletitle{Comparing desktop vs. Mobile interaction for the creation of pervasive augmented reality experiences}.
\newblock \bibinfo{journal}{\emph{Journal of Imaging}} \bibinfo{volume}{8}, \bibinfo{number}{3} (\bibinfo{year}{2022}), \bibinfo{pages}{79}.
\newblock


\bibitem[Mandelbaum(2012)]%
        {mandelbaum2012storytelling}
\bibfield{author}{\bibinfo{person}{Jenny Mandelbaum}.} \bibinfo{year}{2012}\natexlab{}.
\newblock \showarticletitle{Storytelling in conversation}.
\newblock \bibinfo{journal}{\emph{The handbook of conversation analysis}} (\bibinfo{year}{2012}), \bibinfo{pages}{492--507}.
\newblock


\bibitem[Markouzis and Fessakis(2015)]%
        {markouzis2015interactive}
\bibfield{author}{\bibinfo{person}{Dimitrios Markouzis} {and} \bibinfo{person}{Georgios Fessakis}.} \bibinfo{year}{2015}\natexlab{}.
\newblock \showarticletitle{Interactive storytelling and mobile augmented reality applications for learning and entertainment—a rapid prototyping perspective}. In \bibinfo{booktitle}{\emph{2015 International Conference on Interactive Mobile Communication Technologies and Learning (IMCL)}}. IEEE, \bibinfo{pages}{4--8}.
\newblock


\bibitem[MediaPipe({[n.\,d.]})]%
        {mediapipe}
\bibfield{author}{\bibinfo{person}{MediaPipe}.} \bibinfo{year}{[n.\,d.]}\natexlab{}.
\newblock \bibinfo{title}{MediaPipe}.
\newblock \bibinfo{howpublished}{\url{https://mediapipe.dev/}}.
\newblock


\bibitem[Mehri et~al\mbox{.}(2016)]%
        {mehri2016samplernn}
\bibfield{author}{\bibinfo{person}{Soroush Mehri}, \bibinfo{person}{Kundan Kumar}, \bibinfo{person}{Ishaan Gulrajani}, \bibinfo{person}{Rithesh Kumar}, \bibinfo{person}{Shubham Jain}, \bibinfo{person}{Jose Sotelo}, \bibinfo{person}{Aaron Courville}, {and} \bibinfo{person}{Yoshua Bengio}.} \bibinfo{year}{2016}\natexlab{}.
\newblock \showarticletitle{SampleRNN: An unconditional end-to-end neural audio generation model}.
\newblock \bibinfo{journal}{\emph{arXiv preprint arXiv:1612.07837}} (\bibinfo{year}{2016}).
\newblock


\bibitem[Miller(2019)]%
        {miller2019digital}
\bibfield{author}{\bibinfo{person}{Carolyn~Handler Miller}.} \bibinfo{year}{2019}\natexlab{}.
\newblock \bibinfo{booktitle}{\emph{Digital Storytelling 4e: A creator's guide to interactive entertainment}}.
\newblock \bibinfo{publisher}{CRC Press}.
\newblock


\bibitem[Mokady et~al\mbox{.}(2022)]%
        {mokady2022self}
\bibfield{author}{\bibinfo{person}{Ron Mokady}, \bibinfo{person}{Omer Tov}, \bibinfo{person}{Michal Yarom}, \bibinfo{person}{Oran Lang}, \bibinfo{person}{Inbar Mosseri}, \bibinfo{person}{Tali Dekel}, \bibinfo{person}{Daniel Cohen-Or}, {and} \bibinfo{person}{Michal Irani}.} \bibinfo{year}{2022}\natexlab{}.
\newblock \showarticletitle{Self-distilled stylegan: Towards generation from internet photos}. In \bibinfo{booktitle}{\emph{ACM SIGGRAPH 2022 Conference Proceedings}}. \bibinfo{pages}{1--9}.
\newblock


\bibitem[mozilla({[n.\,d.]})]%
        {mozilla}
\bibfield{author}{\bibinfo{person}{mozilla}.} \bibinfo{year}{[n.\,d.]}\natexlab{}.
\newblock \bibinfo{title}{Web Speech API}.
\newblock \bibinfo{howpublished}{\url{https://developer.mozilla.org/en-US/docs/Web/API/Web_Speech_API}}.
\newblock


\bibitem[Ni et~al\mbox{.}(2023)]%
        {ni2023conditional}
\bibfield{author}{\bibinfo{person}{Haomiao Ni}, \bibinfo{person}{Changhao Shi}, \bibinfo{person}{Kai Li}, \bibinfo{person}{Sharon~X Huang}, {and} \bibinfo{person}{Martin~Renqiang Min}.} \bibinfo{year}{2023}\natexlab{}.
\newblock \showarticletitle{Conditional Image-to-Video Generation with Latent Flow Diffusion Models}. In \bibinfo{booktitle}{\emph{Proceedings of the IEEE/CVF Conference on Computer Vision and Pattern Recognition}}. \bibinfo{pages}{18444--18455}.
\newblock


\bibitem[Nichol et~al\mbox{.}(2021)]%
        {nichol2021glide}
\bibfield{author}{\bibinfo{person}{Alex Nichol}, \bibinfo{person}{Prafulla Dhariwal}, \bibinfo{person}{Aditya Ramesh}, \bibinfo{person}{Pranav Shyam}, \bibinfo{person}{Pamela Mishkin}, \bibinfo{person}{Bob McGrew}, \bibinfo{person}{Ilya Sutskever}, {and} \bibinfo{person}{Mark Chen}.} \bibinfo{year}{2021}\natexlab{}.
\newblock \showarticletitle{Glide: Towards photorealistic image generation and editing with text-guided diffusion models}.
\newblock \bibinfo{journal}{\emph{arXiv preprint arXiv:2112.10741}} (\bibinfo{year}{2021}).
\newblock


\bibitem[O'Meara and Szita(2021)]%
        {o2021ar}
\bibfield{author}{\bibinfo{person}{Jennifer O'Meara} {and} \bibinfo{person}{Kata Szita}.} \bibinfo{year}{2021}\natexlab{}.
\newblock \showarticletitle{AR cinema: Visual storytelling and embodied experiences with augmented reality filters and backgrounds}.
\newblock \bibinfo{journal}{\emph{PRESENCE: Virtual and Augmented Reality}}  \bibinfo{volume}{30} (\bibinfo{year}{2021}), \bibinfo{pages}{99--123}.
\newblock


\bibitem[Oord et~al\mbox{.}(2016)]%
        {oord2016wavenet}
\bibfield{author}{\bibinfo{person}{Aaron van~den Oord}, \bibinfo{person}{Sander Dieleman}, \bibinfo{person}{Heiga Zen}, \bibinfo{person}{Karen Simonyan}, \bibinfo{person}{Oriol Vinyals}, \bibinfo{person}{Alex Graves}, \bibinfo{person}{Nal Kalchbrenner}, \bibinfo{person}{Andrew Senior}, {and} \bibinfo{person}{Koray Kavukcuoglu}.} \bibinfo{year}{2016}\natexlab{}.
\newblock \showarticletitle{Wavenet: A generative model for raw audio}.
\newblock \bibinfo{journal}{\emph{arXiv preprint arXiv:1609.03499}} (\bibinfo{year}{2016}).
\newblock


\bibitem[Palombini(2017)]%
        {palombini2017storytelling}
\bibfield{author}{\bibinfo{person}{Augusto Palombini}.} \bibinfo{year}{2017}\natexlab{}.
\newblock \showarticletitle{Storytelling and telling history. Towards a grammar of narratives for Cultural Heritage dissemination in the Digital Era}.
\newblock \bibinfo{journal}{\emph{Journal of cultural heritage}}  \bibinfo{volume}{24} (\bibinfo{year}{2017}), \bibinfo{pages}{134--139}.
\newblock


\bibitem[Park et~al\mbox{.}(2015)]%
        {park2015storytelling}
\bibfield{author}{\bibinfo{person}{Seung-Bo Park}, \bibinfo{person}{Jason~J Jung}, {and} \bibinfo{person}{EunSoon You}.} \bibinfo{year}{2015}\natexlab{}.
\newblock \showarticletitle{Storytelling of collaborative learning system on augmented reality}.
\newblock \bibinfo{journal}{\emph{New Trends in Computational Collective Intelligence}} (\bibinfo{year}{2015}), \bibinfo{pages}{139--147}.
\newblock


\bibitem[Patashnik et~al\mbox{.}(2021)]%
        {patashnik2021styleclip}
\bibfield{author}{\bibinfo{person}{Or Patashnik}, \bibinfo{person}{Zongze Wu}, \bibinfo{person}{Eli Shechtman}, \bibinfo{person}{Daniel Cohen-Or}, {and} \bibinfo{person}{Dani Lischinski}.} \bibinfo{year}{2021}\natexlab{}.
\newblock \showarticletitle{Styleclip: Text-driven manipulation of stylegan imagery}. In \bibinfo{booktitle}{\emph{Proceedings of the IEEE/CVF International Conference on Computer Vision}}. \bibinfo{pages}{2085--2094}.
\newblock


\bibitem[Pavlik and Bridges(2013)]%
        {pavlik2013emergence}
\bibfield{author}{\bibinfo{person}{John~V Pavlik} {and} \bibinfo{person}{Frank Bridges}.} \bibinfo{year}{2013}\natexlab{}.
\newblock \showarticletitle{The emergence of augmented reality (AR) as a storytelling medium in journalism}.
\newblock \bibinfo{journal}{\emph{Journalism \& Communication Monographs}} \bibinfo{volume}{15}, \bibinfo{number}{1} (\bibinfo{year}{2013}), \bibinfo{pages}{4--59}.
\newblock


\bibitem[Peng et~al\mbox{.}(2020)]%
        {peng2020non}
\bibfield{author}{\bibinfo{person}{Kainan Peng}, \bibinfo{person}{Wei Ping}, \bibinfo{person}{Zhao Song}, {and} \bibinfo{person}{Kexin Zhao}.} \bibinfo{year}{2020}\natexlab{}.
\newblock \showarticletitle{Non-autoregressive neural text-to-speech}. In \bibinfo{booktitle}{\emph{International conference on machine learning}}. PMLR, \bibinfo{pages}{7586--7598}.
\newblock


\bibitem[Peterson and Langellier(2006)]%
        {peterson2006communication}
\bibfield{author}{\bibinfo{person}{Eric~E Peterson} {and} \bibinfo{person}{Kristin~M Langellier}.} \bibinfo{year}{2006}\natexlab{}.
\newblock \showarticletitle{Communication as storytelling}.
\newblock \bibinfo{journal}{\emph{Communication as... Perspectives on Theory}} (\bibinfo{year}{2006}), \bibinfo{pages}{123--131}.
\newblock


\bibitem[Petridis et~al\mbox{.}(2023)]%
        {petridis2023anglekindling}
\bibfield{author}{\bibinfo{person}{Savvas Petridis}, \bibinfo{person}{Nicholas Diakopoulos}, \bibinfo{person}{Kevin Crowston}, \bibinfo{person}{Mark Hansen}, \bibinfo{person}{Keren Henderson}, \bibinfo{person}{Stan Jastrzebski}, \bibinfo{person}{Jeffrey~V Nickerson}, {and} \bibinfo{person}{Lydia~B Chilton}.} \bibinfo{year}{2023}\natexlab{}.
\newblock \showarticletitle{Anglekindling: Supporting journalistic angle ideation with large language models}. In \bibinfo{booktitle}{\emph{Proceedings of the 2023 CHI Conference on Human Factors in Computing Systems}}. \bibinfo{pages}{1--16}.
\newblock


\bibitem[Pintar(2023)]%
        {pintar2023invisible}
\bibfield{author}{\bibinfo{person}{Judith Pintar}.} \bibinfo{year}{2023}\natexlab{}.
\newblock \showarticletitle{Invisible, aesthetic, and enrolled listeners across storytelling modalities: Immersive preference as situated player type}.
\newblock \bibinfo{journal}{\emph{Convergence}} (\bibinfo{year}{2023}), \bibinfo{pages}{13548565231206505}.
\newblock


\bibitem[Plappert et~al\mbox{.}(2016)]%
        {plappert2016kit}
\bibfield{author}{\bibinfo{person}{Matthias Plappert}, \bibinfo{person}{Christian Mandery}, {and} \bibinfo{person}{Tamim Asfour}.} \bibinfo{year}{2016}\natexlab{}.
\newblock \showarticletitle{The KIT motion-language dataset}.
\newblock \bibinfo{journal}{\emph{Big data}} \bibinfo{volume}{4}, \bibinfo{number}{4} (\bibinfo{year}{2016}), \bibinfo{pages}{236--252}.
\newblock


\bibitem[Poole et~al\mbox{.}(2022)]%
        {poole2022dreamfusion}
\bibfield{author}{\bibinfo{person}{Ben Poole}, \bibinfo{person}{Ajay Jain}, \bibinfo{person}{Jonathan~T Barron}, {and} \bibinfo{person}{Ben Mildenhall}.} \bibinfo{year}{2022}\natexlab{}.
\newblock \showarticletitle{Dreamfusion: Text-to-3d using 2d diffusion}.
\newblock \bibinfo{journal}{\emph{arXiv preprint arXiv:2209.14988}} (\bibinfo{year}{2022}).
\newblock


\bibitem[Radford et~al\mbox{.}(2021)]%
        {radford2021learning}
\bibfield{author}{\bibinfo{person}{Alec Radford}, \bibinfo{person}{Jong~Wook Kim}, \bibinfo{person}{Chris Hallacy}, \bibinfo{person}{Aditya Ramesh}, \bibinfo{person}{Gabriel Goh}, \bibinfo{person}{Sandhini Agarwal}, \bibinfo{person}{Girish Sastry}, \bibinfo{person}{Amanda Askell}, \bibinfo{person}{Pamela Mishkin}, \bibinfo{person}{Jack Clark}, {et~al\mbox{.}}} \bibinfo{year}{2021}\natexlab{}.
\newblock \showarticletitle{Learning transferable visual models from natural language supervision}. In \bibinfo{booktitle}{\emph{International conference on machine learning}}. PMLR, \bibinfo{pages}{8748--8763}.
\newblock


\bibitem[Raeburn et~al\mbox{.}(2022)]%
        {raeburn2022developing}
\bibfield{author}{\bibinfo{person}{Gideon Raeburn}, \bibinfo{person}{Martin Welton}, {and} \bibinfo{person}{Laurissa Tokarchuk}.} \bibinfo{year}{2022}\natexlab{}.
\newblock \showarticletitle{Developing a play-anywhere handheld AR storytelling app using remote data collection}.
\newblock \bibinfo{journal}{\emph{Frontiers in Computer Science}}  \bibinfo{volume}{4} (\bibinfo{year}{2022}), \bibinfo{pages}{927177}.
\newblock


\bibitem[Ramesh et~al\mbox{.}(2022)]%
        {ramesh2022hierarchical}
\bibfield{author}{\bibinfo{person}{Aditya Ramesh}, \bibinfo{person}{Prafulla Dhariwal}, \bibinfo{person}{Alex Nichol}, \bibinfo{person}{Casey Chu}, {and} \bibinfo{person}{Mark Chen}.} \bibinfo{year}{2022}\natexlab{}.
\newblock \showarticletitle{Hierarchical text-conditional image generation with clip latents}.
\newblock \bibinfo{journal}{\emph{arXiv preprint arXiv:2204.06125}} \bibinfo{volume}{1}, \bibinfo{number}{2} (\bibinfo{year}{2022}), \bibinfo{pages}{3}.
\newblock


\bibitem[resfulapi({[n.\,d.]})]%
        {restful}
\bibfield{author}{\bibinfo{person}{resfulapi}.} \bibinfo{year}{[n.\,d.]}\natexlab{}.
\newblock \bibinfo{title}{RESTFul API}.
\newblock \bibinfo{howpublished}{\url{https://restfulapi.net/}}.
\newblock


\bibitem[Rol{\'o}n-Dow(2011)]%
        {rolon2011race}
\bibfield{author}{\bibinfo{person}{Rosalie Rol{\'o}n-Dow}.} \bibinfo{year}{2011}\natexlab{}.
\newblock \showarticletitle{Race (ing) stories: Digital storytelling as a tool for critical race scholarship}.
\newblock \bibinfo{journal}{\emph{Race Ethnicity and Education}} \bibinfo{volume}{14}, \bibinfo{number}{2} (\bibinfo{year}{2011}), \bibinfo{pages}{159--173}.
\newblock


\bibitem[Rombach et~al\mbox{.}(2022)]%
        {rombach2022high}
\bibfield{author}{\bibinfo{person}{Robin Rombach}, \bibinfo{person}{Andreas Blattmann}, \bibinfo{person}{Dominik Lorenz}, \bibinfo{person}{Patrick Esser}, {and} \bibinfo{person}{Bj{\"o}rn Ommer}.} \bibinfo{year}{2022}\natexlab{}.
\newblock \showarticletitle{High-resolution image synthesis with latent diffusion models}. In \bibinfo{booktitle}{\emph{Proceedings of the IEEE/CVF conference on computer vision and pattern recognition}}. \bibinfo{pages}{10684--10695}.
\newblock


\bibitem[Ruiz et~al\mbox{.}(2023)]%
        {ruiz2023dreambooth}
\bibfield{author}{\bibinfo{person}{Nataniel Ruiz}, \bibinfo{person}{Yuanzhen Li}, \bibinfo{person}{Varun Jampani}, \bibinfo{person}{Yael Pritch}, \bibinfo{person}{Michael Rubinstein}, {and} \bibinfo{person}{Kfir Aberman}.} \bibinfo{year}{2023}\natexlab{}.
\newblock \showarticletitle{Dreambooth: Fine tuning text-to-image diffusion models for subject-driven generation}. In \bibinfo{booktitle}{\emph{Proceedings of the IEEE/CVF Conference on Computer Vision and Pattern Recognition}}. \bibinfo{pages}{22500--22510}.
\newblock


\bibitem[Saharia et~al\mbox{.}(2022)]%
        {saharia2022photorealistic}
\bibfield{author}{\bibinfo{person}{Chitwan Saharia}, \bibinfo{person}{William Chan}, \bibinfo{person}{Saurabh Saxena}, \bibinfo{person}{Lala Li}, \bibinfo{person}{Jay Whang}, \bibinfo{person}{Emily~L Denton}, \bibinfo{person}{Kamyar Ghasemipour}, \bibinfo{person}{Raphael Gontijo~Lopes}, \bibinfo{person}{Burcu Karagol~Ayan}, \bibinfo{person}{Tim Salimans}, {et~al\mbox{.}}} \bibinfo{year}{2022}\natexlab{}.
\newblock \showarticletitle{Photorealistic text-to-image diffusion models with deep language understanding}.
\newblock \bibinfo{journal}{\emph{Advances in Neural Information Processing Systems}}  \bibinfo{volume}{35} (\bibinfo{year}{2022}), \bibinfo{pages}{36479--36494}.
\newblock


\bibitem[Salo et~al\mbox{.}(2016)]%
        {salo2016backend}
\bibfield{author}{\bibinfo{person}{Kari Salo}, \bibinfo{person}{Diana Giova}, {and} \bibinfo{person}{Tommi Mikkonen}.} \bibinfo{year}{2016}\natexlab{}.
\newblock \showarticletitle{Backend infrastructure supporting audio augmented reality and storytelling}. In \bibinfo{booktitle}{\emph{Human Interface and the Management of Information: Applications and Services: 18th International Conference, HCI International 2016 Toronto, Canada, July 17-22, 2016. Proceedings, Part II 18}}. Springer, \bibinfo{pages}{325--335}.
\newblock


\bibitem[Schegloff(1997)]%
        {schegloff1997narrative}
\bibfield{author}{\bibinfo{person}{Emanuel~A Schegloff}.} \bibinfo{year}{1997}\natexlab{}.
\newblock \showarticletitle{" Narrative analysis" thirty years later}.
\newblock \bibinfo{journal}{\emph{Journal of narrative and life history}} \bibinfo{volume}{7}, \bibinfo{number}{1-4} (\bibinfo{year}{1997}), \bibinfo{pages}{97--106}.
\newblock


\bibitem[Shi et~al\mbox{.}(2025)]%
        {shi2025caring}
\bibfield{author}{\bibinfo{person}{Jingyu Shi}, \bibinfo{person}{Rahul Jain}, \bibinfo{person}{Seunggeun Chi}, \bibinfo{person}{Hyungjun Doh}, \bibinfo{person}{Hyung-gun Chi}, \bibinfo{person}{Alexander~J Quinn}, {and} \bibinfo{person}{Karthik Ramani}.} \bibinfo{year}{2025}\natexlab{}.
\newblock \showarticletitle{CARING-AI: Towards Authoring Context-aware Augmented Reality INstruction through Generative Artificial Intelligence}. In \bibinfo{booktitle}{\emph{Proceedings of the 2025 CHI Conference on Human Factors in Computing Systems}}. \bibinfo{pages}{1--23}.
\newblock


\bibitem[Shi et~al\mbox{.}(2023a)]%
        {shi2023hci}
\bibfield{author}{\bibinfo{person}{Jingyu Shi}, \bibinfo{person}{Rahul Jain}, \bibinfo{person}{Hyungjun Doh}, \bibinfo{person}{Ryo Suzuki}, {and} \bibinfo{person}{Karthik Ramani}.} \bibinfo{year}{2023}\natexlab{a}.
\newblock \showarticletitle{An HCI-centric survey and taxonomy of human-generative-AI interactions}.
\newblock \bibinfo{journal}{\emph{arXiv preprint arXiv:2310.07127}} (\bibinfo{year}{2023}).
\newblock


\bibitem[Shi et~al\mbox{.}(2023b)]%
        {shi2023understanding}
\bibfield{author}{\bibinfo{person}{Jingyu Shi}, \bibinfo{person}{Rahul Jain}, \bibinfo{person}{Runlin Duan}, {and} \bibinfo{person}{Karthik Ramani}.} \bibinfo{year}{2023}\natexlab{b}.
\newblock \showarticletitle{Understanding Generative AI in Art: An Interview Study with Artists on G-AI from an HCI Perspective}.
\newblock \bibinfo{journal}{\emph{arXiv preprint arXiv:2310.13149}} (\bibinfo{year}{2023}).
\newblock


\bibitem[Shin et~al\mbox{.}(2022)]%
        {shin2022effects}
\bibfield{author}{\bibinfo{person}{Jae-eun Shin}, \bibinfo{person}{Boram Yoon}, \bibinfo{person}{Dooyoung Kim}, {and} \bibinfo{person}{Woontack Woo}.} \bibinfo{year}{2022}\natexlab{}.
\newblock \showarticletitle{The Effects of Spatial Complexity on Narrative Experience in Space-Adaptive AR Storytelling}.
\newblock \bibinfo{journal}{\emph{IEEE Transactions on Visualization and Computer Graphics}} (\bibinfo{year}{2022}).
\newblock


\bibitem[{\c{S}}im{\c{s}}ek and Direk{\c{c}}i(2023)]%
        {csimcsek2023effects}
\bibfield{author}{\bibinfo{person}{Bilal {\c{S}}im{\c{s}}ek} {and} \bibinfo{person}{Bekir Direk{\c{c}}i}.} \bibinfo{year}{2023}\natexlab{}.
\newblock \showarticletitle{The effects of augmented reality storybooks on student's reading comprehension}.
\newblock \bibinfo{journal}{\emph{British Journal of Educational Technology}} \bibinfo{volume}{54}, \bibinfo{number}{3} (\bibinfo{year}{2023}), \bibinfo{pages}{754--772}.
\newblock


\bibitem[Singer et~al\mbox{.}(2022)]%
        {singer2022make}
\bibfield{author}{\bibinfo{person}{Uriel Singer}, \bibinfo{person}{Adam Polyak}, \bibinfo{person}{Thomas Hayes}, \bibinfo{person}{Xi Yin}, \bibinfo{person}{Jie An}, \bibinfo{person}{Songyang Zhang}, \bibinfo{person}{Qiyuan Hu}, \bibinfo{person}{Harry Yang}, \bibinfo{person}{Oron Ashual}, \bibinfo{person}{Oran Gafni}, {et~al\mbox{.}}} \bibinfo{year}{2022}\natexlab{}.
\newblock \showarticletitle{Make-a-video: Text-to-video generation without text-video data}.
\newblock \bibinfo{journal}{\emph{arXiv preprint arXiv:2209.14792}} (\bibinfo{year}{2022}).
\newblock


\bibitem[Singh et~al\mbox{.}(2021)]%
        {singh2021story}
\bibfield{author}{\bibinfo{person}{Abbey Singh}, \bibinfo{person}{Ramanpreet Kaur}, \bibinfo{person}{Peter Haltner}, \bibinfo{person}{Matthew Peachey}, \bibinfo{person}{Mar Gonzalez-Franco}, \bibinfo{person}{Joseph Malloch}, {and} \bibinfo{person}{Derek Reilly}.} \bibinfo{year}{2021}\natexlab{}.
\newblock \showarticletitle{Story creatar: a toolkit for spatially-adaptive augmented reality storytelling}. In \bibinfo{booktitle}{\emph{2021 IEEE Virtual Reality and 3D User Interfaces (VR)}}. IEEE, \bibinfo{pages}{713--722}.
\newblock


\bibitem[Stylianidou et~al\mbox{.}(2020)]%
        {stylianidou2020helping}
\bibfield{author}{\bibinfo{person}{Nayia Stylianidou}, \bibinfo{person}{Angelos Sofianidis}, \bibinfo{person}{Elpiniki Manoli}, {and} \bibinfo{person}{Maria Meletiou-Mavrotheris}.} \bibinfo{year}{2020}\natexlab{}.
\newblock \showarticletitle{“Helping Nemo!”—Using augmented reality and alternate reality games in the context of universal design for learning}.
\newblock \bibinfo{journal}{\emph{Education Sciences}} \bibinfo{volume}{10}, \bibinfo{number}{4} (\bibinfo{year}{2020}), \bibinfo{pages}{95}.
\newblock


\bibitem[Taylor et~al\mbox{.}(2018)]%
        {taylor2018digital}
\bibfield{author}{\bibinfo{person}{Murray Taylor}, \bibinfo{person}{Mauricio Marrone}, \bibinfo{person}{Mark Tayar}, {and} \bibinfo{person}{Beate Mueller}.} \bibinfo{year}{2018}\natexlab{}.
\newblock \showarticletitle{Digital storytelling and visual metaphor in lectures: a study of student engagement}.
\newblock \bibinfo{journal}{\emph{Accounting Education}} \bibinfo{volume}{27}, \bibinfo{number}{6} (\bibinfo{year}{2018}), \bibinfo{pages}{552--569}.
\newblock


\bibitem[Tevet et~al\mbox{.}(2022a)]%
        {tevet2022motionclip}
\bibfield{author}{\bibinfo{person}{Guy Tevet}, \bibinfo{person}{Brian Gordon}, \bibinfo{person}{Amir Hertz}, \bibinfo{person}{Amit~H Bermano}, {and} \bibinfo{person}{Daniel Cohen-Or}.} \bibinfo{year}{2022}\natexlab{a}.
\newblock \showarticletitle{Motionclip: Exposing human motion generation to clip space}. In \bibinfo{booktitle}{\emph{European Conference on Computer Vision}}. Springer, \bibinfo{pages}{358--374}.
\newblock


\bibitem[Tevet et~al\mbox{.}(2022b)]%
        {tevet2022human}
\bibfield{author}{\bibinfo{person}{Guy Tevet}, \bibinfo{person}{Sigal Raab}, \bibinfo{person}{Brian Gordon}, \bibinfo{person}{Yonatan Shafir}, \bibinfo{person}{Daniel Cohen-Or}, {and} \bibinfo{person}{Amit~H Bermano}.} \bibinfo{year}{2022}\natexlab{b}.
\newblock \showarticletitle{Human motion diffusion model}.
\newblock \bibinfo{journal}{\emph{arXiv preprint arXiv:2209.14916}} (\bibinfo{year}{2022}).
\newblock


\bibitem[Tyurina(2023)]%
        {tyurina2023leveraging}
\bibfield{author}{\bibinfo{person}{Anastasia Tyurina}.} \bibinfo{year}{2023}\natexlab{}.
\newblock \showarticletitle{Leveraging AR-Driven Visual Storytelling to Enhance Communication of Complex Social Issues: Principles, Strategies, and Multifaceted Roles of Visuals}. In \bibinfo{booktitle}{\emph{SIGGRAPH Asia 2023 Educator's Forum}}. \bibinfo{pages}{1--2}.
\newblock


\bibitem[Van~Laer et~al\mbox{.}(2019)]%
        {van2019storytelling}
\bibfield{author}{\bibinfo{person}{Tom Van~Laer}, \bibinfo{person}{Stephanie Feiereisen}, {and} \bibinfo{person}{Luca~M Visconti}.} \bibinfo{year}{2019}\natexlab{}.
\newblock \showarticletitle{Storytelling in the digital era: A meta-analysis of relevant moderators of the narrative transportation effect}.
\newblock \bibinfo{journal}{\emph{Journal of Business Research}}  \bibinfo{volume}{96} (\bibinfo{year}{2019}), \bibinfo{pages}{135--146}.
\newblock


\bibitem[Vera and S{\'a}nchez(2016)]%
        {vera2016model}
\bibfield{author}{\bibinfo{person}{Fernando Vera} {and} \bibinfo{person}{J~Alfredo S{\'a}nchez}.} \bibinfo{year}{2016}\natexlab{}.
\newblock \showarticletitle{A model for in-situ augmented reality content creation based on storytelling and gamification}. In \bibinfo{booktitle}{\emph{Proceedings of the 6th Mexican Conference on Human-Computer Interaction}}. \bibinfo{pages}{39--42}.
\newblock


\bibitem[Wachowiak and Gromann(2023)]%
        {wachowiak2023does}
\bibfield{author}{\bibinfo{person}{Lennart Wachowiak} {and} \bibinfo{person}{Dagmar Gromann}.} \bibinfo{year}{2023}\natexlab{}.
\newblock \showarticletitle{Does GPT-3 Grasp Metaphors? Identifying Metaphor Mappings with Generative Language Models}. In \bibinfo{booktitle}{\emph{Proceedings of the 61st Annual Meeting of the Association for Computational Linguistics (Volume 1: Long Papers)}}. \bibinfo{pages}{1018--1032}.
\newblock


\bibitem[Walwema(2015)]%
        {walwema2015art}
\bibfield{author}{\bibinfo{person}{Josephine Walwema}.} \bibinfo{year}{2015}\natexlab{}.
\newblock \showarticletitle{The Art of Storytelling.}
\newblock \bibinfo{journal}{\emph{Writing \& Pedagogy}} \bibinfo{volume}{7}, \bibinfo{number}{1} (\bibinfo{year}{2015}).
\newblock


\bibitem[Wang et~al\mbox{.}(2021)]%
        {wang2021gesturar}
\bibfield{author}{\bibinfo{person}{Tianyi Wang}, \bibinfo{person}{Xun Qian}, \bibinfo{person}{Fengming He}, \bibinfo{person}{Xiyun Hu}, \bibinfo{person}{Yuanzhi Cao}, {and} \bibinfo{person}{Karthik Ramani}.} \bibinfo{year}{2021}\natexlab{}.
\newblock \showarticletitle{Gesturar: An authoring system for creating freehand interactive augmented reality applications}. In \bibinfo{booktitle}{\emph{The 34th Annual ACM Symposium on User Interface Software and Technology}}. \bibinfo{pages}{552--567}.
\newblock


\bibitem[Yamamoto et~al\mbox{.}(2020)]%
        {yamamoto2020parallel}
\bibfield{author}{\bibinfo{person}{Ryuichi Yamamoto}, \bibinfo{person}{Eunwoo Song}, {and} \bibinfo{person}{Jae-Min Kim}.} \bibinfo{year}{2020}\natexlab{}.
\newblock \showarticletitle{Parallel WaveGAN: A fast waveform generation model based on generative adversarial networks with multi-resolution spectrogram}. In \bibinfo{booktitle}{\emph{ICASSP 2020-2020 IEEE International Conference on Acoustics, Speech and Signal Processing (ICASSP)}}. IEEE, \bibinfo{pages}{6199--6203}.
\newblock


\bibitem[Ye et~al\mbox{.}(2020)]%
        {ye2020aranimator}
\bibfield{author}{\bibinfo{person}{Hui Ye}, \bibinfo{person}{Kin~Chung Kwan}, \bibinfo{person}{Wanchao Su}, {and} \bibinfo{person}{Hongbo Fu}.} \bibinfo{year}{2020}\natexlab{}.
\newblock \showarticletitle{ARAnimator: In-situ character animation in mobile AR with user-defined motion gestures}.
\newblock \bibinfo{journal}{\emph{ACM Transactions on Graphics (TOG)}} \bibinfo{volume}{39}, \bibinfo{number}{4} (\bibinfo{year}{2020}), \bibinfo{pages}{83--1}.
\newblock


\bibitem[Yilmaz and Goktas(2017)]%
        {yilmaz2017using}
\bibfield{author}{\bibinfo{person}{Rabia~Meryem Yilmaz} {and} \bibinfo{person}{Yuksel Goktas}.} \bibinfo{year}{2017}\natexlab{}.
\newblock \showarticletitle{Using augmented reality technology in storytelling activities: examining elementary students’ narrative skill and creativity}.
\newblock \bibinfo{journal}{\emph{Virtual Reality}}  \bibinfo{volume}{21} (\bibinfo{year}{2017}), \bibinfo{pages}{75--89}.
\newblock


\bibitem[Yu et~al\mbox{.}(2022)]%
        {yu2022scaling}
\bibfield{author}{\bibinfo{person}{Jiahui Yu}, \bibinfo{person}{Yuanzhong Xu}, \bibinfo{person}{Jing~Yu Koh}, \bibinfo{person}{Thang Luong}, \bibinfo{person}{Gunjan Baid}, \bibinfo{person}{Zirui Wang}, \bibinfo{person}{Vijay Vasudevan}, \bibinfo{person}{Alexander Ku}, \bibinfo{person}{Yinfei Yang}, \bibinfo{person}{Burcu~Karagol Ayan}, {et~al\mbox{.}}} \bibinfo{year}{2022}\natexlab{}.
\newblock \showarticletitle{Scaling autoregressive models for content-rich text-to-image generation}.
\newblock \bibinfo{journal}{\emph{arXiv preprint arXiv:2206.10789}} \bibinfo{volume}{2}, \bibinfo{number}{3} (\bibinfo{year}{2022}), \bibinfo{pages}{5}.
\newblock


\bibitem[Zhang et~al\mbox{.}(2022)]%
        {zhang2022motiondiffuse}
\bibfield{author}{\bibinfo{person}{Mingyuan Zhang}, \bibinfo{person}{Zhongang Cai}, \bibinfo{person}{Liang Pan}, \bibinfo{person}{Fangzhou Hong}, \bibinfo{person}{Xinying Guo}, \bibinfo{person}{Lei Yang}, {and} \bibinfo{person}{Ziwei Liu}.} \bibinfo{year}{2022}\natexlab{}.
\newblock \showarticletitle{Motiondiffuse: Text-driven human motion generation with diffusion model}.
\newblock \bibinfo{journal}{\emph{arXiv preprint arXiv:2208.15001}} (\bibinfo{year}{2022}).
\newblock


\bibitem[Zhao and Ma(2018)]%
        {zhao2018compensation}
\bibfield{author}{\bibinfo{person}{Zhenjie Zhao} {and} \bibinfo{person}{Xiaojuan Ma}.} \bibinfo{year}{2018}\natexlab{}.
\newblock \showarticletitle{A compensation method of two-stage image generation for human-ai collaborated in-situ fashion design in augmented reality environment}. In \bibinfo{booktitle}{\emph{2018 IEEE International Conference on Artificial Intelligence and Virtual Reality (AIVR)}}. IEEE, \bibinfo{pages}{76--83}.
\newblock


\bibitem[Zhou et~al\mbox{.}(2004)]%
        {zhou2004interactive}
\bibfield{author}{\bibinfo{person}{ZhiYing Zhou}, \bibinfo{person}{Adrian~David Cheok}, \bibinfo{person}{JiunHorng Pan}, {and} \bibinfo{person}{Yu Li}.} \bibinfo{year}{2004}\natexlab{}.
\newblock \showarticletitle{An interactive 3D exploration narrative interface for storytelling}. In \bibinfo{booktitle}{\emph{Proceedings of the 2004 conference on Interaction design and children: building a community}}. \bibinfo{pages}{155--156}.
\newblock


\end{thebibliography}

\appendix
\section{Stories}
\label{apdx:stories}
In this section, we manifest the five stories we used in our study.
We marked four atomic elements to augment in color, namely {\color{teal}Characters}, {\color{blue}Backgrounds}, {\color{magenta}Sentiments}, and {\color{violet}Development}.
\subsection{Story - 1}
Once in {\color{blue}a small, mountain-ringed village}, lived Emily, {\color{teal}a young painter} whose heart {\color{magenta}brimmed with passion} but also {\color{magenta}frustration}. She loved {\color{blue}the natural world}, yet felt her paintings never truly captured its essence.

One day, Emily {\color{violet}stumbled upon} {\color{teal}a hidden waterfall} in the forest, {\color{magenta}a scene so breathtaking} it seemed otherworldly. {\color{magenta}Inspired yet challenged}, she set out to paint it. Her usual methods felt inadequate against the waterfall's majesty. She {\color{violet}spent days and nights experimenting} with new styles and techniques, striving to capture not just the sight, but {\color{magenta}the emotion it stirred within} her.

Through this tireless effort, Emily realized her art shouldn't just replicate a scene; it needed to convey the feelings it evoked. Finally, she {\color{violet}created {\color{teal}a masterpiece painting}}, vastly different from her previous works. It was {\color{magenta}vibrant, alive}, {\color{teal}a visual echo of the waterfall's spirit}.

When exhibited, her painting moved {\color{teal}the villagers}, evoking the sensation of being right there at the waterfall. Emily's journey transformed her artistic style and understanding of art itself. She learned that true art was about conveying emotion and connecting deeply with others. Her new approach brought not just {\color{magenta}satisfaction}, but {\color{magenta}recognition}, and {\color{violet}she continued exploring and capturing {\color{blue}the world's beauty}}, infusing her emotions into her canvases.

\subsection{Story - 2}
I've always been drawn to the stars. Ever since I was {\color{teal}a child}, lying on the grass, gazing up at {\color{teal}the night sky}, I felt a connection to {\color{blue}the cosmos}. My name is Alex, and I am {\color{teal}an astronomer}.

My life revolves around {\color{teal}telescopes} and celestial charts, seeking to unlock the mysteries of the universe. But it was {\color{blue}one particular night} that changed everything for me. {\color{violet}As I peered through my telescope}, I discovered a celestial phenomenon - {\color{teal}a star that seemed to pulsate with an unusual light}. This discovery consumed me, and {\color{violet}I dedicated months to studying it}, often forgetting the world around me. The more I studied, the more I realized that this star's behavior defied all our known laws of astrophysics.

This revelation was not just a scientific breakthrough; it made me question everything we knew about the universe. It humbled me, reminding me of the infinite {\color{blue}mysteries of space}. {\color{violet}Sharing my findings with the world} brought new {\color{magenta}excitement} to {\color{teal}the astronomical community}, and together, we embarked on a journey of redefining our understanding of the cosmos.

\subsection{Story - 3}
{\color{blue}On a remote coastline}, battered by relentless waves, stood {\color{teal}an old lighthouse}, it lights {\color{teal}a solitary beacon} {\color{blue}in the foggy nights}. The lighthouse keeper, {\color{teal}an old seafarer} named Captain Holloway, {\color{violet}spent his days in \color{magenta}}, tending to the lighthouse and reminiscing about his long-lost love, {\color{teal}a woman as enigmatic and wild as the sea itself}. Their love was a {\color{magenta}tempest, intense and fleeting}, as she disappeared {\color{blue}one stormy night}, leaving Holloway with only memories. Each evening, he would play {\color{teal}a {\color{magenta}melancholic} melody} on his {\color{teal}violin}, the notes carried by the wind, hoping they would reach her wherever she might be.

{\color{blue}One stormy night}, as Holloway played his violin, a figure appeared in the rain - it was her, drawn back by the sorrowful {\color{teal}melody}. {\color{violet}They shared one last night together}, their love rekindled amidst {\color{teal}the storm}. By morning, {\color{violet}she was gone again}, but this time, Holloway felt {\color{magenta}a sense of closure}. {\color{violet}He continued to tend the lighthouse}, his heart no longer heavy with loss, but filled with gratitude for the brief reunion granted by the sea's mysterious ways.

\subsection{Story - 4}
{\color{blue}Deep in the heart of the dense jungle}, hidden under layers of time and nature, lay {\color{blue}the ruins of an ancient city}, once a bustling center of {\color{blue}an advanced civilization}, now forgotten by the world. Its {\color{teal}sole guardian} was an ageless being known as Arion, {\color{teal}a creature both human and spirit}, bound to the city by {\color{teal}an ancient curse}. Arion's existence was a {\color{magenta}solitary} one, his only purpose {\color{violet}to protect {\color{teal}the secrets and artifacts of the lost city}}. For centuries, he kept {\color{teal}intruders} at bay, using his powers to create illusions and manipulate the jungle's treacherous paths.

His resolve was unyielding until {\color{violet}the arrival of {\color{teal}a team of explorers}}, led by {\color{teal}a determined archaeologist named Dr. Elena Mendoza}. Unlike others before, they sought not treasure, but knowledge. Moved by their genuine quest for understanding, Arion revealed himself and the city's history. {\color{violet}He guided them through the ruins}, each structure and artifact a testament to the city's {\color{teal}once-great civilization}. With Arion's help, Dr. Mendoza and her team were able to {\color{violet}unravel the mysteries of the city}, bringing {\color{teal}its forgotten history} to light. Arion, having fulfilled his purpose, was {\color{magenta}finally released} from his curse, {\color{violet}allowing the spirit of the lost city to live on} through the pages of history.

\subsection{Story - 5}
My name is Isabella, and I am {\color{teal}an artist}, though the world has long forgotten me. I used to paint {\color{teal}vibrant landscapes and portraits} full of emotion, but {\color{violet}as the years passed, my art fell into obscurity}.

Living in {\color{blue}a small, quiet town}, my days were spent walking through {\color{blue}the streets}, my heart heavy with unspoken stories and unseen art. One day, {\color{violet}while wandering, I stumbled upon} {\color{teal}an abandoned building}. Inside, I found {\color{teal}walls covered in graffiti} - a tapestry of untold stories. This discovery {\color{magenta}sparked a fire within} me. I began to paint again, but this time, {\color{violet}I painted on those walls}. My art became a fusion of my traditional style and the raw, expressive nature of graffiti. As I {\color{violet}transformed the walls into {\color{teal}a gallery of emotions}}, people started to visit, drawn by the {\color{magenta}curious blend of art} styles. My forgotten art found a new audience, and more importantly, I rediscovered my {\color{magenta}passion}.

Through this unexpected journey, I learned that art is not just about recognition, but about {\color{magenta}the joy of creation} and the impact it has on others, even in the most unexpected places.

\end{document}